\documentclass[preprint,prd,aps,showpacs,showkeys,nofootinbib]{revtex4}
\usepackage{epsfig}
\usepackage{graphicx}
\usepackage{amsmath}
\usepackage{amssymb}
\usepackage{booktabs}
\usepackage{xcolor}
\usepackage{float}
\usepackage{dcolumn}
\usepackage{hyperref}

\begin{document}

\title{Electric dipole moments from the perspective of a scalar triplet and singlet extension of the MSSM: A study of neutrons, electrons, mercury, and b and c quarks}

\author{Qing-hua Li$^{1,2}$\footnote{lqhlqh202408.163.com},Jin-Lei Yang$^{1,2,3}$\footnote{jlyang@hbu.edu.cn},Xiang Yang$^{1,2,3}$\footnote{yangxiang202406@163.com},
	Tai-Fu Feng$^{1,2,3,4}$\footnote{fengtf@hbu.edu.cn}}

\affiliation{$^1$ Department of Physics, Hebei University, Baoding 071002, China}
\affiliation{$^2$ Hebei Key Laboratory of High-precision Computation and Application of Quantum Field Theory, Baoding, 071002, China}
\affiliation{$^3$ Hebei Research Center of the Basic Discipline for Computational Physics, Baoding, 071002, China}
\date{\today}
\affiliation{$^4$ Department of Physics, Chongqing University, Chongqing 401331, China}

	\begin{abstract}
In the framework of the minimal supersymmetric model extension with new scalar triplets and singlet (TNMSSM), we analyze the electric dipole moment (EDM) of neutron ($d_n$), electron EDM($d_e$), the  mercury EDM($d_{Hg}$), $b$ quark ($d_b$) and $c$ quark ($d_c$) by considering the contributions from the one-loop diagrams, some two-loop diagrams and the Weinberg operators. The effects of TNMSSM specific CPV sources $\chi_d,\;\chi_t$ on $d_n$, $d_e$, $d_{Hg}$, $d_b$, $d_c$ are specialized, it is found that they have significant contributions to these EDMs, and the current upper bounds on $d_n$ impose strict constraints on $\chi_d,\;\chi_t$. The theoretical predictions on $d_b$, $d_c$ can reach about $10^{-22}$ e$\cdot$cm and $10^{-23}$ e$\cdot$cm respectively by taking the upper bounds on $d_n$, $d_e$, $d_{Hg}$ into account, which have great potential to be observed in future.
		
	\end{abstract}

	\maketitle
	
	\section{Introduction\label{sec1}}
In the early universe, the interaction of high-energy particles would lead to the symmetric production of matter and antimatter. However, the observations of baryon asymmetry in the current universe indicate this symmetry was broken under some mechanism~\cite{Abe:2001xe}. This mechanism requires more CP-violating (CPV) sources beyond the standard model (SM)~\cite{Modak:2021vre,Aubert:2002ic}, because the sole sources of CPV in the SM are the Cabbibo-Kobayashi-Maskawa (CKM) phases at quark sector, and the computational results indicate that the produced values are too small to adequately account for the observed asymmetry between matter and antimatter~\cite{Dine:2003,Modak:2020uyq,Masip:1998za}. The introducing of new CPV phases may make significant contributions to the electric dipole moments (EDMs) of particles, while these EDMs are highly suppressed in the SM~\cite{Yang:2019aao}. Hence, the observations of the EDMs can provide clear signals of CPV effects in NP~\cite{Ellis:1982tk,Polchinski:1983zd,Nath:1991dn,Kizukuri:1992nj,Kizukuri:1991mb}.

In this work, we focus on the EDMs of neutron ($d_n$), electron EDM($d_e$), the  mercury EDM($d_{Hg}$), $b$ quark ($d_b$) and $c$ quark ($d_c$) in the framework of the minimal supersymmetric model (MSSM) extension with new scalar triplets and singlet (TNMSSM)~\cite{2Y0,3Y0,King:2015oxa}. Although $d_n$, $d_e$, $d_{Hg}$ $d_b$, $d_c$ are not observed experimentally so far, strict upper bounds on them have been obtained~\cite{PDG2023,de-1,de-3,Blinov:2008mu,ParticleDataGroup2024,Graner:2016ses,Sala:2013osa}
\begin{eqnarray}
	&&|d_n|<1.8\times10^{-26}{\rm e\cdot cm},\nonumber\\
	&&|d_e|<4.1\times10^{-30}{\rm e\cdot cm},\nonumber\\
	&&|d_{Hg}|<7.4\times10^{-30}{\rm e\cdot cm},\nonumber\\
	&&|d_b|<2.0\times10^{-17}{\rm e\cdot cm},\nonumber\\
	&&|d_c|<4.4\times10^{-17}{\rm e\cdot cm}.		
	\label{EDMlimits}
\end{eqnarray}
In addition, the chromoelectric dipole moments (CEDM) of heavy quarks is tightly constrained by the upper bound on $d_n$, which can be expressed as~\cite{Pospelov:2000bw}
\begin{eqnarray}
	&&d_n=(1\pm0.5)[1.4(d_d^\gamma-0.25d_u^\gamma)+1.1e(d_d^g+0.5d_u^g)]\pm(22\pm10){\;\rm MeV}C_5,
	\label{EDMdn}
\end{eqnarray}
$d_q^\gamma$, $d_q^g$, $C_5$ represent the EDM of quark $q$ from the electroweak interaction, CEDM of quark $q$ and the coefficient of the Weinberg operator at the chirality scale, respectively. Using the running from Refs.~\cite{Braaten:1990gq,Degrassi:2005zd}, $d_{d,u}^{\gamma,g}$ and $C_5$ can be expressed in the form of $d_c^g$ at the $m_c$ scale~\cite{Sala:2013osa}. Subsequently, a new upper limit $|d_c^g|<1.0\times10^{-22}{\rm cm}$ can be obtained by considering the limits on $d_n$. Through the calculation of the rigorous constraints on $d_b^g$ in Refs.~\cite{Chang:1990jv,Gisbert:2019ftm,Sala:2013osa}, a new limits for the EDMs of $b$ and $c$ quarks can be derived
\begin{eqnarray}
	&&|d_b|<1.2\times10^{-20}{\rm e\cdot cm},\nonumber\\
	&&|d_c|<1.5\times10^{-21}{\rm e\cdot cm}.
	\label{EDMbc}
\end{eqnarray}
The above results improve the previous bounds in Eq.~(\ref{EDMlimits}) by about three orders of magnitude. It is obvious that the experimental upper limits on these quantities are strict, hence the contributions from new CPV phases in NP models may be tightly constrained, and researching NP effects on these EDMs may help to elucidate the mechanism of CPV~\cite{Espinosa:1992hp,Espinosa:1998re}.

The TNMSSM extends the MSSM with two $SU(2)_L$ triplets and an additional scalar singlet~\cite{5}. The $\mu$ problem of the MSSM~\cite{NMSSM1,NMSSM2} can be solved by introducing the scalar singlet in the TNMSSM, and $\mu$ is at about the electroweak scale naturally after the scalar singlet achieving nonzero vacuum expectation value (VEV)~\cite{Espinosa:2015kwx, Espinosa:2011ax, Elias-Miro:2011sqh}. The $\mu$ term is the primary source of baryon asymmetry, and large phase of $\mu$ required by baryon asymmetry enhances the theoretical predictions on the EDMs $d_n$, $d_b$, $d_c$, $d_e$, $d_{Hg}$ significantly. Similarly, mutual cancellations between different phases are most likely to suppress the EDM to below the corresponding experimental upper limits \cite{Falk:1996ni,Falk:1998pu,Brhlik:1998zn,Bartl:1999bc,Abel:2001vy,Barger:2001nu,Olive:2005ru,Cirigliano:2006dg}. And the so-called little hierarchy problem can also be alleviated by introducing two scalar triplets. Correspondingly, there are also new CPV sources in the TNMSSM compared with the MSSM, hence we explore the effects of CPV sources in the TNMSSM on $d_n$, $d_b$, $d_c$, $d_e$, $d_{Hg}$ in this work.

The paper is organized as follows. In Section II, the overview of the key elements of the TNMSSM are briefly summarized by introducing the superpotential and the general soft breaking terms. Then Section III analyzes the EDM of neutron $d_n$, $b$ quark $d_b$, $c$ quark $d_c$, electron EDM $d_e$ and mercury EDM $d_{Hg}$. In order to see the corrections to these EDMs clearly, the numerical results of $d_n$, $d_b$, $d_c$, $d_e$, $d_{Hg}$ with new CPV phases are explored in Section IV. Conclusions are summarized in Section V.	

	\section{The TNMSSM\label{sec2}}
Besides the superfields $\hat{H_d}$, $\hat{H_u}$ of the MSSM, TNMSSM introduces a guage singlet superfield S and two $SU(2)_L$ triplet superfields $T$ and $\bar{T} $~\cite{2Y0,3Y0}
\begin{eqnarray}
	&&\hat{T}=\left(\begin{array}{cc}\frac{1}{\sqrt2}T^+,&-T^{++}\\ T^0,&\frac{-1}{\sqrt2}T^+\end{array}\right)\sim (1, 3, 1),\quad\; \hat{\bar T}=\left(\begin{array}{cc}\frac{1}{\sqrt2}{\bar T}^-,&-{\bar T}^0\\ {\bar T}^{--},&\frac{-1}{\sqrt2}{\bar T}^-\end{array}\right)\sim (-1, 3, 1), \nonumber\\
	&&\hat{H_d}=\left(\begin{array}{c}H_d^0\\ H_d^-\end{array}\right)\sim (-1/2, 2, 1),\quad\;\hat{H_u}=\left(\begin{array}{c}H_u^+\\ H_u^0\end{array}\right)\sim(1/2, 2, 1),\quad\;\hat S \sim(0, 1, 1).
\end{eqnarray}
$T^+$, $\bar T^-$, $H_d^-$, $H_u^-$ are singly-charged Higgs, $T^{++}$, $\bar T^{--}$ are doubly-charged Higgs, $H_d^0$, $H_u^0$, $T^0$, ${\bar T}^0$, $S$ are complex neutral superfields ~\cite{YaserAyazi:2006zw,Barger:2008wn,FileviezPerez:2008sx}
\begin{eqnarray}
	&&H_d^0=\frac{1}{\sqrt2} (v_d +{\rm Re}{H_d^0} +i {\rm Im}{H_d^0}),\quad\; H_u^0=\frac{1}{\sqrt2} (v_u +{\rm Re} {H_u^0} +i{\rm Im}{H_u^0}), \nonumber\\
	&&T^0=\frac{1}{\sqrt2} (v_T +{\rm Re}{T^0} +i{\rm Im}{T^0}),\quad\; {\bar T}^0=\frac{1}{\sqrt2} (v_{\bar T} +{\rm Re}{{\bar T}^0} +i{\rm Im}{{\bar T}^0}), \nonumber\\
	&&S=\frac{1}{\sqrt2} (v_S +{\rm Re} S +i{\rm Im}S),
\end{eqnarray}
where $v_d$, $v_u$, $v_T$, $v_{\bar T}$ and $v_S$ are the corresponding VEVs.

The superpotential of the TNMSSM can be represented as ~\cite{Chen:2023zfl}
\begin{eqnarray}
	&&W_{\rm TNMSSM}=Y_u \hat U^c \hat H_u \cdot \hat Q -Y_d \hat D^c \hat H_d \cdot \hat Q -Y_e \hat E^c \hat H_d \cdot \hat L+\chi_d \hat H_d \cdot \hat T \hat H_d \nonumber\\
	&&\qquad\qquad\quad\;\;+\chi_u \hat H_u \cdot \hat {\bar T} \hat H_u +\frac{1}{3}\kappa \hat S \hat S \hat S +\lambda \hat S \hat H_u \cdot \hat H_d +\Lambda_T \hat S \mathrm{Tr}(\hat {\bar T} \hat T).
\end{eqnarray}
The soft SUSY breaking terms are~\cite{Hu:2024slu}
\begin{eqnarray}
	&&\mathcal{-L}_{\rm soft}=m_{H_u}^2 |H_u|^2 +m_{H_d}^2 |H_d|^2 +m_S^2 |S|^2 +m_{T}^2 \mathrm{Tr}(|T|^2) +m_{\bar T}^2 \mathrm{Tr}(|\bar T|^2)
	\nonumber\\
	&&\hspace{1.4cm}
	+m_{Q}^2 |Q|^2 +m_{u}^2 |u|^2 +m_{d}^2 |d|^2 +(T_{\Lambda_T} S \mathrm{Tr}(T \bar T) +T_{\lambda} S H_u \cdot H_d +\frac{1}{3} T_{\kappa} S^3
	\nonumber\\
	&&\hspace{1.4cm}
	-T_{\chi_u} H_u \cdot \bar T H_u -T_{\chi_d} H_d \cdot T H_d + T_{u,ij} \tilde{Q_j} \cdot H_u \tilde{u^c_i} -T_{d,ij} \tilde{Q_j} \cdot H_d \tilde{d^c_i} +H.c.),
\end{eqnarray}
where~\cite{Zhang:2024ijy}
\begin{eqnarray}
	&&H_u \cdot H_d =H_u^+ H_d^- -H_u^0 H_d^0, \nonumber\\
	&&H_d \cdot T H_d=\sqrt{2} H_d^- H_d^0 T^+ -(H_d^0)^2 T^0 -(H_d^-)^2 T^{++}, \nonumber\\
	&&H_u \cdot \bar T H_u=\sqrt{2} H_u^+ H_u^0 {\bar T}^- -(H_u^0)^2 {\bar T}^0 -(H_u^+)^2 {\bar T}^{--}.
\end{eqnarray}

	\section{EDMs in the TNMSSM\label{sec3}}
In this section, we present the EDMs of b, c quark $d_b$, $d_c$, neutron $d_n$, mercury $d_{Hg}$ and electron $d_e$.
	
	\subsection{The EDMs of b, c quark $d_b$, $d_c$\label{sec3.1}}
The quark EDMs at the low scale $\Lambda_{\chi}$, can be obtained from $d_{q}^\gamma(\Lambda_{\chi})$, $d_{q}^g(\Lambda_{\chi})$ and $C_{5}(\Lambda_{\chi})$ by
\begin{eqnarray}
	&&d_q=d_q^\gamma(\Lambda_\chi)+\frac{e}{4\pi}d_q^g(\Lambda_\chi)+\frac{e\Lambda_\chi}{4\pi}C_5(\Lambda_\chi),
	\label{eq11}
\end{eqnarray}
where $\Lambda_\chi=m_q$ denotes the chirality breaking scale, $m_q$ denotes the corresponding quark mass. Weinberg discovered the supersymmetric contribution of the purely gluonic CP-violating operator to the neutron EDM, the Wilson coefficient $C_5$ reads \cite{EDM2}
\begin{eqnarray}
	&&C_5(\Lambda)=-\frac{3g_3^5}{(4\pi)^4M_3^3}\Big\{m_t \Im[e^{2i\theta_3}(Z_{\tilde t})_{2,2}(Z_{\tilde t})_{2,1}^\dagger]\frac{x_{\tilde t_1}-x_{\tilde t_2}}{x_{M_3}}H\Big(\frac{x_{\tilde t_1}}{x_{M_3}},\frac{x_{\tilde t_2}}{x_{M_3}},\frac{x_t}{x_{M_3}}\Big)\nonumber\\
	&&\qquad\qquad\;\;+m_b \Im[e^{2i\theta_3}(Z_{\tilde b})_{2,2}(Z_{\tilde b})_{2,1}^\dagger]\frac{x_{\tilde b_1}-x_{\tilde b_2}}{x_{M_3}}H\Big(\frac{x_{\tilde b_1}}{x_{M_3}},\frac{x_{\tilde b_2}}{x_{M_3}},\frac{x_b}{x_{M_3}}\Big)\Big\}.
	\label{eq12}
\end{eqnarray}
Here, the $\Im[x,y]$ refers to the imaginary part of $[x,y]$, the $H$ function can be found in Refs.~\cite{EDM2,EDM3}. $ Z_{{\tilde t}}$ and $ Z_{{\tilde b}}$ are respectively the matrix to diagonalize the mass squared matrix of stop and sbottom. At the same time, the evolution of $d_q^\gamma$, $d_q^g$ and $C_5$ with the renormalization group equations from the matching scale $\Lambda$ to the chiral symmetry breaking scale $\Lambda_\chi$ is represented by
\begin{eqnarray}
	&&d_q^\gamma(\Lambda_\chi)=\eta^{ED}d_q^\gamma(\Lambda),\;d_q^g(\Lambda_\chi)=\eta^{CD}d_q^g(\Lambda),\;
	C_5(\Lambda_\chi)=\eta^GC_5(\Lambda).
	\label{eq13}
\end{eqnarray}
where $\eta^{ED}$, $\eta^{CD}$, $\eta^G$ represent the evolution coefficient of $d_q^\gamma$, $d_q^g$ and $C_5$ with the renormalization group equations from the higher scale $\Lambda$ to the low scale $\Lambda_\chi$.

The effective Lagrangian for the EDM $d_{q}$ of the fermion is defined through the dimension five operator \cite{EDM1}.
\begin{eqnarray}
	&&\mathcal{L}_{EDM}=-\frac{i}{2}d_q^\gamma\bar q \sigma^{\mu\nu}\gamma_5 q F_{\mu\nu},
	\label{eq1}
\end{eqnarray}
with $\sigma^{\mu\nu}=i[\gamma^\mu,\gamma^\nu]/2$, $F_{{\mu\nu}}$ representing the electromagnetic field strength, $q$ denoting a fermion field. Besides the operator in Eq.~(\ref{eq1}), the CEDM of quarks can also contributes to $d_q$
\begin{eqnarray}
	&&\mathcal{L}_{CEDM}=-\frac{i}{2} d_q^g \bar q\sigma^{\mu\nu}\gamma_5 q G_{\mu\nu}^a T^a,
\end{eqnarray}
where $G_{\mu\nu}$ is the $SU(3)$ gauge field strength, $T^a$ is the $SU(3)$ generators.

We use the effective method to obtain the effective Lagrangian with the CPV operators at matching scale $\Lambda$ which should be evolved down to the quark mass scale using the renormalization group equations (RGEs). The effective Lagrangian with these CPV operators relevant to the quark EDM and CEDM can be represented by Eq.~(\ref{eq16})
\begin{eqnarray}
	&&{\cal L}_{{eff}}=\sum\limits_{i}^5C_{i}(\Lambda){\cal O}_{i}(\Lambda)\;,\
	\label{eq16}
\end{eqnarray}
with
\begin{eqnarray}
	&&{\cal O}_{1}=\overline{q}\sigma^{\mu\nu}P_{L}qF_{{\mu\nu}}
	\;,~~~~~{\cal O}_{2}=\overline{q}\sigma^{\mu\nu}P_{R}qF_{{\mu\nu}}
	\;,~~~{\cal O}_{3}=\overline{q}T^a\sigma^{\mu\nu}P_{L}qG^a_{{\mu\nu}}
	\;,\nonumber\\
	&&{\cal O}_{4}=\overline{q}T^a\sigma^{\mu\nu}P_{R}qG^a_{{\mu\nu}}
	\;,~{\cal O}_{5}=-{1\over6}f_{{abc}}G_{{\mu\rho}}^aG^{b\rho}_{\nu}
	G_{{\lambda\sigma}}^c\epsilon^{\mu\nu\lambda\sigma}\;.
	\label{eq17}
\end{eqnarray}
$C_{i}(\Lambda)$ are the Wilson coefficients, like $C_{5}$. The effective Lagrangian of quarks and squarks with gluino, neutralino and chargino are mentioned in Appendix~\ref{Lagrangian}~\cite{Cao:2014kya,Ibrahim:2004gb}.

The purely gluonic Weinberg operator $O_{5}$ originates from the two-loop "gluino-squark" diagrams. The results obtained at the matching scale $\Lambda$ should be transformed down to the chirality breaking scale $\Lambda_{\chi}$. After caculation, the following relations for the coefficients are obtained \cite{EDM6}
\begin{eqnarray}
	&&C_5(\Lambda_{\chi})=\mathcal{K}^{\gamma_{GG}/\beta} C_5(\Lambda),~~~
	C_\gamma(\Lambda_{\chi})=\mathcal{K}^{\gamma_{q}/\beta} C_\gamma(\Lambda),\nonumber\\&&
	C_q(\Lambda_{\chi})=\mathcal{K}^{\gamma_{qq}/\beta}C_q(\Lambda)
	+C_{5}(\Lambda)\frac{\gamma_{Gq}m_q(\Lambda)}{\gamma_{qq}+\gamma_m-\gamma_{GG}}(\mathcal{K}^{\gamma_{qq}/\beta}-\mathcal{K}^{(\gamma_{qq}-\gamma_m)/\beta}),
\end{eqnarray}
with \cite{EDM6}

\begin{eqnarray}
	&& \mathcal{K}=\frac{g_s(\mu)}{g_s(M)}, ~~~\gamma(\mathcal{O}_q)=\gamma_{qq}=\frac{29-2N_f}{3},~~~\gamma_q=\frac{8}{3},
	\nonumber\\&&\gamma_{GG}=-3-2N_f, ~~~\gamma_{Gq}=6, ~~~\gamma_m=-8,~~~
	\beta=\frac{33-2N_f}{3},\label{chz4}
\end{eqnarray}
where $N_f$ is the number of light quarks at scale $\Lambda_{\chi}$. From this, the term in Eq.~(\ref{eq11}) can be further calculated using the following $C_\gamma(\Lambda),~ C_q(\Lambda)$ and $ C_5(\Lambda)$ \cite{EDM7}
\begin{eqnarray}
	&&d_{q}^\gamma=C_\gamma(\Lambda_{\chi})=\eta^{ED}C_\gamma(\Lambda),~~~~
	d_{q}^g=C_q(\Lambda_{\chi})=\eta^{CD}C_q(\Lambda),\nonumber\\&&
	C_{5}(\Lambda_{\chi})=\eta^GC_5(\Lambda).
\end{eqnarray}

Thus, it is possible to calculate the evolution coefficient in Eq.(\ref{eq13})
\begin{eqnarray}
	&&d_q^\gamma(\Lambda_\chi)=1/1.53d_q^\gamma(\Lambda),\;d_q^g(\Lambda_\chi)=3.3d_q^g(\Lambda),\;
	C_5(\Lambda_\chi)=3.3C_5(\Lambda).
	\label{19}
\end{eqnarray}
The EDMs of quark can be written as
\begin{eqnarray}
	&&d_q^\gamma=-\frac{2e Q_q m_q}{(4\pi)^2}\Im(C_2^R+C_2^{L*}+C_6^R),
	\label{EDM}
\end{eqnarray}
where $C_{2,6}^{L,R}$ represent the Wilson coefficients of the corresponding operators $O_{2,6}^{L,R}$, it is expressed as follows:
\begin{eqnarray}
	&&O_2^{L,R}=\frac{e Q_q}{(4\pi)^2}(-iD_\alpha^*) \bar q \gamma^\alpha F\cdot\sigma P_{L,R}q,\nonumber\\
	&&O_6^{L,R}=\frac{e Q_q m_q}{(4\pi)^2}\bar q F\cdot\sigma P_{L,R}q.
\end{eqnarray}

Similarly, the quark CEDMs can be written as
\begin{eqnarray}
	&&d_q^g=-\frac{2g_3m_q}{(4\pi)^2}\Im(C_7^R+C_7^{L*}+C_8^R),
	\label{CEDM}
\end{eqnarray}
where $C_{7,8}^{L,R}$ represent the Wilson coefficients of the corresponding operators $O_{7,8}^{L,R}$
\begin{eqnarray}
	&&O_7^{L,R}=\frac{g_3}{(4\pi)^2}(-iD_\alpha^*) \bar q \gamma^\alpha G^a\cdot\sigma T^a P_{L,R}q,\nonumber\\
	&&O_8^{L,R}=\frac{g_3m_q}{(4\pi)^2}\bar q G^a\cdot\sigma T^a P_{L,R}q.
\end{eqnarray}

\begin{figure}
	\centering
	\includegraphics[width=0.7\linewidth]{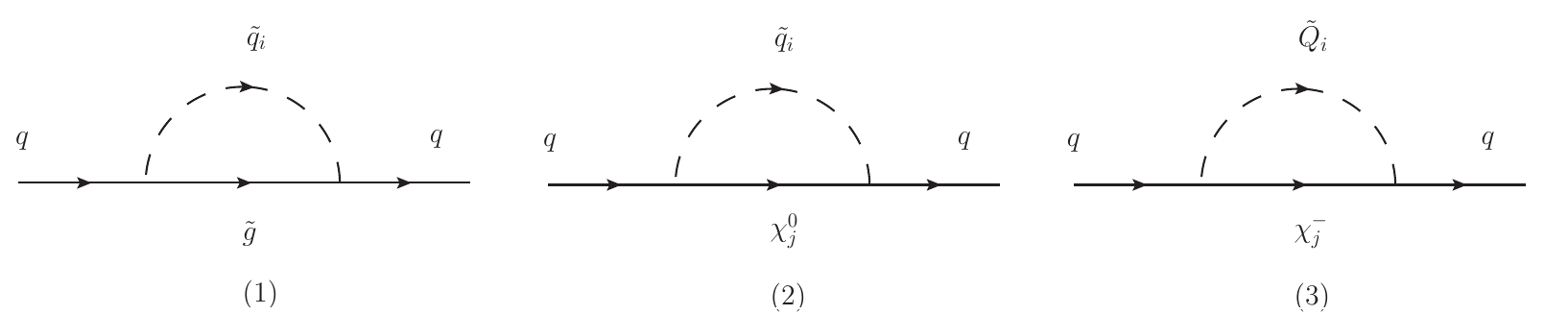}
	\caption{The one-loop diagrams which contributes to $d_q^\gamma$ and $d_q^g$ are obtained by attaching a photon and a gluon respectively to the internal particles in all possible ways.}
	\label{fig:feynmandiagramq}
\end{figure}

The one-loop Feynman diagrams contributing to the above amplitudes are depicted in Fig.~\ref{fig:feynmandiagramq}. Calculating the Feynman diagrams, $d_q^\gamma$ and $d_q^g$ at the one-loop level can be written as
\begin{eqnarray}
	&&d_q^{\gamma(1)}=\frac{e_q e}{12\pi^2m_W}\frac{\sqrt {x_{\tilde g}}}{x_{\tilde q_i}}\Im\Big[C_{\bar{\tilde g} \tilde q_i q}^L C_{\bar q \tilde q_i \tilde g}^L\Big]I_1\Big(\frac{x_{\tilde g}}{x_{\tilde q_i}}\Big),\nonumber\\
	&&d_q^{g(1)}=\frac{-g_3}{32\pi^2m_W}\frac{\sqrt {x_{\tilde g}}}{x_{\tilde q_i}}\Im\Big[C_{\bar {\tilde g} \tilde q_i q}^L C_{\bar q \tilde q_i \tilde g}^L\Big]I_2\Big(\frac{x_{\tilde g}}{x_{\tilde q_i}}\Big),\nonumber\\
	&&d_q^{\gamma(2)}=\frac{e_q e}{32\pi^2m_W}\frac{\sqrt {x_{\chi_j^0}}}{x_{\tilde q_i}}\Im\Big[C_{\bar q \tilde q_i\chi_j^0}^L C_{\bar{\chi}_j^0 \tilde q_i q}^R\Big]I_1\Big(\frac{x_{\chi_j^0}}{x_{\tilde q_i}}\Big),\nonumber\\
	&&d_q^{g(2)}=\frac{g_3^3}{128\pi^2 e^2 m_W}\frac{\sqrt {x_{\chi_j^0}}}{x_{\tilde q_i}}\Im\Big[C_{\bar q \tilde q_i\chi_j^0}^L C_{\bar{\chi}_j^0 \tilde q_i q}^R\Big]I_1\Big(\frac{x_{\chi_j^0}}{x_{\tilde q_i}}\Big),\nonumber\\
	&&d_q^{\gamma(3)}=\frac{e}{16\pi^2m_W}\frac{\sqrt {x_{\chi_j^-}}}{x_{\tilde Q_i}}\Im\Big[C_{\bar q \tilde Q_i\chi_j^-}^L C_{\bar{\chi}_j^- \tilde Q_i q}^R\Big]\Big[e_QI_1\Big(\frac{x_{\chi_j^-}}{x_{\tilde Q_i}}\Big)+(e_q-e_Q)I_3\Big(\frac{x_{\chi_j^-}}{x_{\tilde q_i}}\Big)\Big],\nonumber\\
	&&d_q^{g(3)}=\frac{g_3^3}{16\pi^2e^2m_W}\frac{\sqrt {x_{\chi_j^-}}}{x_{\tilde Q_i}}\Im\Big[C_{\bar q \tilde Q_i\chi_j^-}^L C_{\bar{\chi}_j^- \tilde Q_i q}^R\Big]I_1\Big(\frac{x_{\chi_j^-}}{x_{\tilde Q_i}}\Big),
	\label{17}
\end{eqnarray}
where $x_i$ denotes $m_i^2/m_W^2$, $g_3$ is the strong coupling constant, $C_{abc}^{L,R}$ denotes the constant parts of the interaction vertex about $abc$ which can be obtained through SARAH ~\cite{SAR1,SAR2,SAR3,SAR4,SAR5}, and $a$, $b$, $c$ denote the interacting particles. We will provide the functions $I_{1,2,3}$ in Appendix~\ref{constants}.

\begin{figure}
	\setlength{\unitlength}{1mm}
	\centering
	\includegraphics[width=6in]{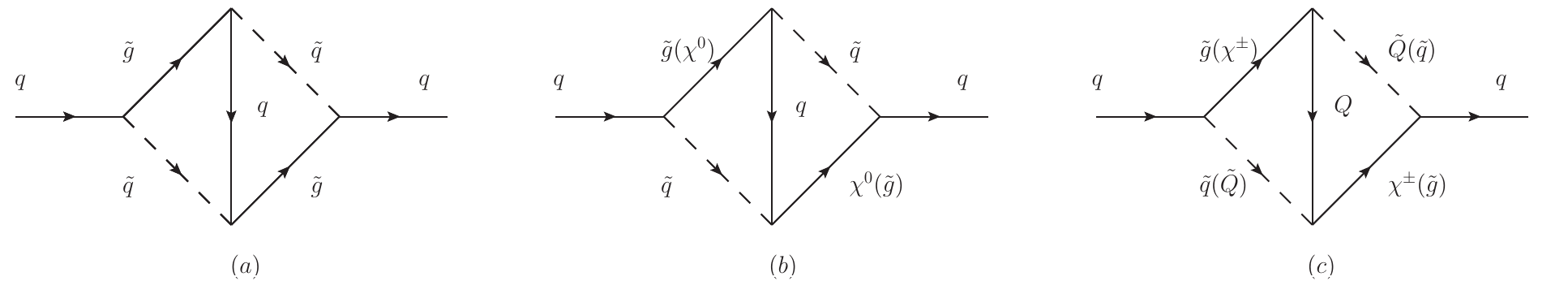}
	\vspace{0cm}
	\caption[]{The two-loop diagrams which contributes to $d_q^\gamma$ and $d_q^g$ are obtained by attaching a photon and a gluon respectively to the internal particles in all possible ways.}
	\label{two-loop Feynman diagram Q}
\end{figure}

The two-loop gluino corrections to the Wilson coefficients from the self-energy diagrams for quarks are considered, the corresponding Feynman diagrams are depicted in Fig.~\ref{two-loop Feynman diagram Q}. The corresponding diploe moment diagrams are obtained by attaching a photon or gluon to the internal particles in all possible ways.
Then the contributions from these two-loop diagrams to $d_q^\gamma$ and $d_q^g$ can be written as
\begin{eqnarray}
	&&d_q^{\gamma(a)}=\frac{-4e_q e g_3^2|m_{\tilde g}|}{9(4\pi)^4m_W^2}F_3(x_{_q},x_{_{\tilde q_j}},x_{_{\tilde g}},x_{_{\tilde g}},x_{_{\tilde q_i}})\Im[C_{\bar q \tilde g \tilde q_j}^L C_{\bar {\tilde g}q\tilde q_j}^L],\nonumber\\
	&&d_q^{g(a)}=d_q^{\gamma(a)}g_3/(e_q e),\nonumber\\
	&&d_q^{\gamma(b)}=\frac{4 e_q e}{3(4\pi)^4m_W^2}\Big\{|m_{\tilde g}|F_4(x_{_q},x_{_{\tilde q_j}},x_{_{\tilde g}},x_{_{\chi_k^0}},x_{_{\tilde q_i}})\Im[C_{\bar \chi_k^0 q \tilde q_j}^RC_{\bar \chi_k^0 q \tilde q_i}^L C_{\bar {\tilde g}q\tilde q_j}^LC_{\bar {\tilde g}q\tilde q_i}^L-C_{\bar \chi_k^0 q \tilde q_j}^L\nonumber\\
	&&\qquad\quad \times C_{\bar \chi_k^0 q \tilde q_i}^RC_{\bar q \tilde g \tilde q_j}^{L*}C_{\bar q \tilde g \tilde q_i}^{L*}]-m_{\chi_k^0}F_5(x_{_q},x_{_{\tilde q_j}},x_{_{\tilde g}},x_{_{\chi_k^0}},x_{_{\tilde q_i}})\Im[C_{\bar \chi_k^0 q \tilde q_j}^RC_{\bar \chi_k^0 q \tilde q_i}^RC_{\bar q \tilde g \tilde q_j}^{L*}C_{\bar {\tilde g}q\tilde q_i}^L\nonumber\\
	&&\qquad\quad-C_{\bar \chi_k^0 q \tilde q_j}^LC_{\bar \chi_k^0 q \tilde q_i}^LC_{\bar q \tilde g \tilde q_i}^{L*}C_{\bar {\tilde g}q\tilde q_j}^L]\Big\},\nonumber\\
	&&d_q^{g(b)}=d_q^{\gamma(b)}g_3/(e_q e),\nonumber\\
	&&d_q^{\gamma(c)}=\frac{2 e}{3(4\pi)^4m_W^2}\Big\{|m_{\tilde g}|F_4(x_{_Q},x_{_{\tilde Q_j}},x_{_{\tilde g}},x_{_{\chi_k^\pm}},x_{_{\tilde q_i}})\Im[C_{\bar Q \chi_k^\pm \tilde q_j}^LC_{\bar q\chi_k^\pm \tilde Q_i}^RC_{\bar{\tilde g}Q\tilde Q_j}^LC_{\bar{\tilde g}q\tilde q_i}^L-C_{\bar Q \chi_k^\pm \tilde q_j}^R\nonumber\\
	&&\qquad\quad\times C_{\bar q\chi_k^\pm \tilde Q_i}^LC_{\bar Q\tilde g\tilde Q_j}^{L*}C_{\bar q\tilde g\tilde q_i}^{L*}]-m_{\chi_k^\pm}F_5(x_{_Q},x_{_{\tilde Q_j}},x_{_{\tilde g}},x_{_{\chi_k^\pm}},x_{_{\tilde q_i}})\Im[C_{\bar Q \chi_k^\pm \tilde q_j}^LC_{\bar q\chi_k^\pm \tilde Q_i}^LC_{\bar Q\tilde g\tilde Q_j}^{L*}C_{\bar{\tilde g}q\tilde q_i}^L\nonumber\\
	&&\qquad\quad-C_{\bar Q \chi_k^\pm \tilde q_j}^RC_{\bar q\chi_k^\pm \tilde Q_i}^RC_{\bar{\tilde g}Q\tilde Q_j}^LC_{\bar q\tilde g\tilde q_i}^{L*}]\Big\},\nonumber\\
	&&d_q^{g(c)}=d_q^{\gamma(c)}g_3/e,
	\label{21}
\end{eqnarray}
where the concrete expressions for the functions $F_{3,4,5}$ can be found in Ref.~\cite{Feng:2004vu}.

\begin{figure}
	\setlength{\unitlength}{1mm}
	\centering
	\includegraphics[width=4in]{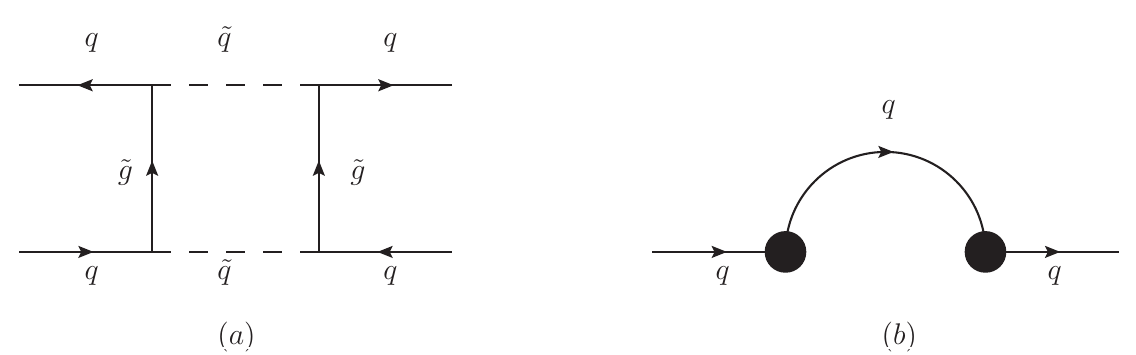}
	\vspace{0cm}
	\caption[]{Full theory diagram (a) and effective diagram (b) are plotted, where the blobs denote the effective vertexes, and a outgoing photon or gluon is attached by all possible ways.}
	\label{matching Feynman diagram}
\end{figure}

We should note that there are infrared divergencies in Fig.~\ref{two-loop Feynman diagram Q} when the SM quarks appear as internal particles, because we calculate these diagrams by expanding the external momentum. In this case, matching full theory diagrams to the corresponding two-loop diagrams in Fig.~\ref{two-loop Feynman diagram Q} is needed to cancel the infrared divergency. Taking Fig.~\ref{two-loop Feynman diagram Q}(a) as example to illustrate how to cancel the infrared divergency, the corresponding diagrams are shown in Fig.~\ref{matching Feynman diagram}.
When the external gluon is attached to an internal particle in Fig.~\ref{two-loop Feynman diagram Q}(a), and the external gluon can be attached to the same internal particle in Fig.~\ref{matching Feynman diagram}(a) or (b). Then infrared divergency in the diagram by attaching a gluon in Fig.~\ref{two-loop Feynman diagram Q}(a) can be cancelled by subtracting the corresponding diagram by attaching a gluon in the same way in Fig.~\ref{matching Feynman diagram}.

\begin{figure}
	\setlength{\unitlength}{1mm}
	\centering
	\includegraphics[width=6in]{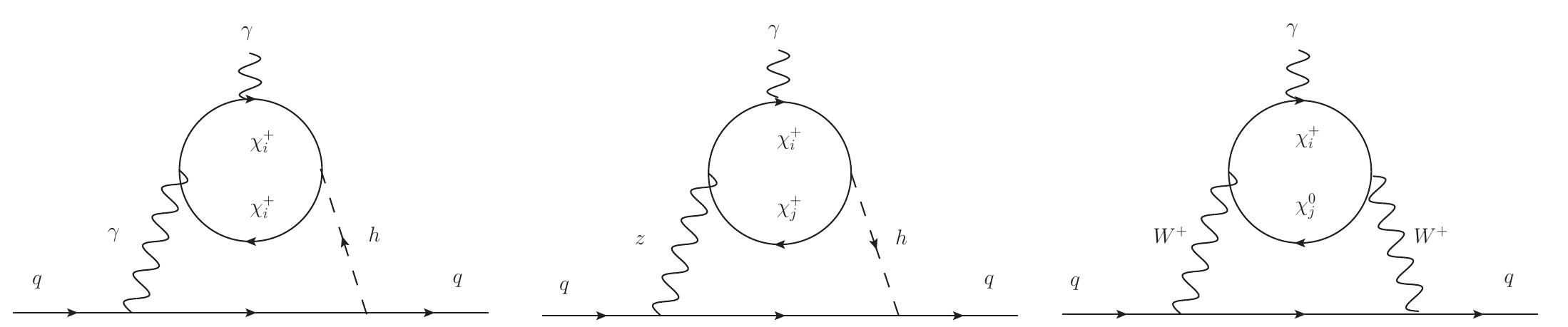}
	\vspace{0cm}
	\caption[]{The two-loop Barr-Zee type diagrams contributing to the quark EDM. The diagrams in which the photon or gluon is emitted from the $W$ boson or the internal fermion do not contribute to the quark EDM or CEDM.}
	\label{two-loop Feynman Bazz}
\end{figure}

In addition, the two-loop Barr-Zee type diagrams can also make contributions to the quark EDM. The diagrams in which a closed fermion loop is attached to the virtual gauge bosons or Higgs fields are considered, and the corresponding Feynman diagrams are depicted in Fig.~\ref{two-loop Feynman Bazz}.
Then, the contributions from these two-loop Barr-Zee type diagrams to $d_q^\gamma$ are given by~\cite{Giudice:2005rz}
\begin{eqnarray}
	&&d_q^{\gamma h}=\frac{e_q e^3}{32\pi^4 m_W}\frac{\sqrt{x_{\chi_i^+}}}{x_{h_k}}\Im\Big[C_{\bar\chi_i^+ h \chi_i^+}^R C_{\bar q h_k q}\Big]f_{\gamma H}\Big(\frac{x_{\chi_i^+}}{x_{h_k}}\Big),\nonumber\\
	&&d_q^{Zh}=\frac{e^2(T_{3q}-2e_q s_w^2)}{128\pi^4 c_w s_w m_W}\frac{\sqrt{x_{\chi_i^+}}}{x_{h_k}}\Im\Big[\Big(C_{\bar\chi_j^+ h_k \chi_i^+}^RC_{\bar\chi_i^+ Z \chi_j^+}^L-C_{\bar\chi_j^+ h_k \chi_i^+}^LC_{\bar\chi_i^+ Z \chi_j^+}^R\Big)C_{\bar q h_k q}\Big]\nonumber\\
	&&\qquad\;\;\;*f_{ZH}\Big(\frac{x_Z}{x_{h_k}},\frac{x_{\chi_i^+}}{x_{h_k}},\frac{x_{\chi_j^+}}{x_{h_k}}\Big),\nonumber\\
	&&d_q^{WW}=\frac{T_{3q} e^3}{128\pi^4 s_w^2 m_W}\sqrt{x_q x_{\chi_i^+}x_{\chi_j^0}}\Im\Big[C_{\bar\chi_j^0 W^+_\mu \chi_i^+}^RC_{\bar\chi_j^0 W^+_\mu \chi_i^+}^{L*}\Big]f_{WW}\Big(x_{\chi_i^+},x_{\chi_j^0}\Big),
	\label{22}
\end{eqnarray}
where $s_w\equiv\sin \theta_W,\;c_w\equiv\cos \theta_W$, and $\theta_W$ is the Weinberg angle, $T_{3q}$ denotes the isospin of the corresponding quark, the functions $f_{\gamma H},\;f_{ZH},\;f_{WW}$ can be found in Ref.~\cite{Giudice:2005rz}.

	\subsection{The EDM of neutron $d_n$ and mercury $d_{Hg}$\label{secA}}

To be consistent with the discussion in Ref.~\cite{Sala:2013osa}, for the neutron EMD $d_n$, we adopt the values $0.5$ and $12{\;\rm MeV}$ for the coefficients $1\pm0.5$ and $22\pm10{\;\rm MeV}$ in Eq.~(\ref{EDMdn}) respectively.
 $C_5$ in Eq.~(\ref{EDMdn}) can be obtain in Eq.(\ref{eq13}) and Eq.(\ref{19}).
  $d_q^g$ and $d_q^\gamma$ in Eq.~(\ref{EDMdn}) have three contributions: from one-loop Feynman diagrams contributions in Fig.~\ref{fig:feynmandiagramq}, which can be obtain in Eq.(\ref{17}), from two-loop Feynman diagrams contributions in Fig.~\ref{two-loop Feynman diagram Q}, which can be obtain in Eq.(\ref{21}) and from two-loop Barr-Zee type diagrams contributions in Fig.~\ref{two-loop Feynman Bazz}, which can be obtain in Eq.(\ref{22}).

According to the Ref.~\cite{Lee:2004we}, where it was shown that the dominant contribution arises from the CEDMs of quarks ($d_q^g$) according to
\begin{eqnarray}
	d_{Hg} = - \left(d_d^g - d_u^g - 0.012 d_s^g \right)
	\times 3.2 \cdot 10^{-2} e  \ .
\end{eqnarray}
$d_q^g$ can be obtained in Eq.(\ref{eq13}), Eq.(\ref{17}) and Eq.(\ref{22}), where $x_i=m_i^2/m_W^2$, $C_{abc}^{L,R}$ denotes the constant parts of the interaction vertex about $abc$ which can be obtained through SARAH ~\cite{SAR1,SAR2,SAR3,SAR4,SAR5}, and $a$, $b$, $c$ denote the interacting particles. We will provide the functions $I_{1,2,3}$ in Appendix~\ref{constants}.

	\subsection{The EDM of electron $d_e$\label{secC}}

	The effective Lagrangian for the electron EDM can be written as
	\begin{eqnarray}
		&&\mathcal{L}_{EDM}=-\frac{i}{2}d_e\bar l_e\sigma^{\mu\nu}\gamma_5 l_e F_{\mu\nu},\label{30}
	\end{eqnarray}
	where $\sigma^{\mu\nu}=i[\gamma^\mu,\gamma^\nu]/2$, $F_{\mu\nu}$ is the electromagnetic field strength, and $l_e$ denotes the electron which is on-shell.
	The electron EDMs can be written as
	\begin{eqnarray}
		&&d_e=-\frac{2eQ_fm_e}{(4\pi)^2}\Im(C_2^R+C_2^{L*}+C_6^R),
		\label{EDMe}
	\end{eqnarray}
	where $C_{2,6}^{L,R}$ represent the Wilson coefficients of the corresponding operators $O_{2,6}^{L,R}$, it is expressed as follows:
	\begin{eqnarray}
		&&O_2^{L,R} = \frac{e Q_q}{(4\pi)^2}(-iD_\alpha^*) \bar l_e  \gamma^\mu F\cdot \sigma P_{L,R} l_e, \nonumber\\
		&&O_6^{L,R} = {eQ_{_q}m_e\over(4\pi)^2} \bar l_e F\cdot \sigma P_{L,R} l_e,
		\label{operators}
	\end{eqnarray}
\begin{figure}
	\setlength{\unitlength}{1mm}
	\centering
	\includegraphics[width=5in]{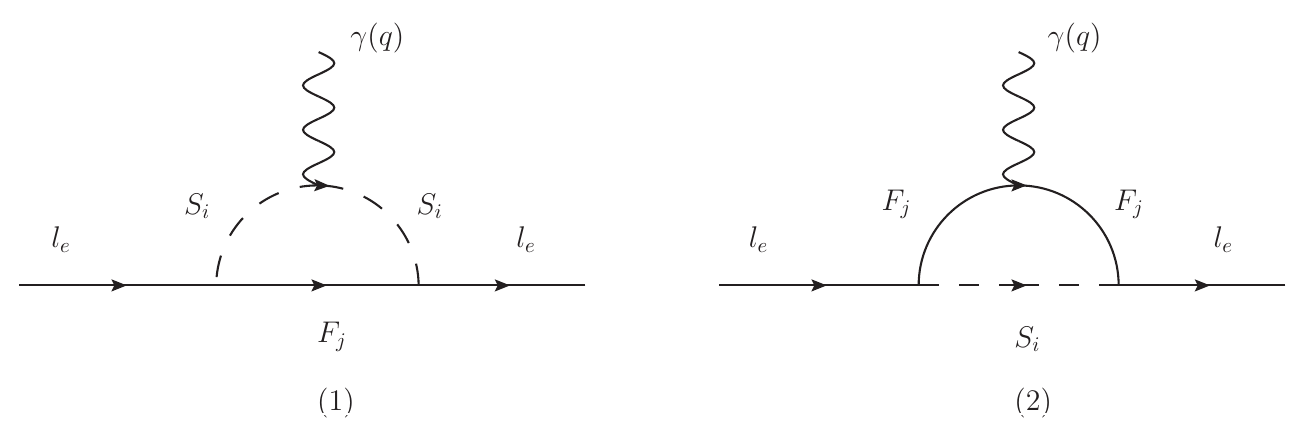}
	\vspace{0cm}
	\caption[]{The one-loop Feynman diagrams contributing to the electron EDM, where (1) denotes a charged scalar loop, and (2) denotes a charged fermion loop.}
	\label{Feynman diagram}
\end{figure}
The one-loop Feynman diagrams contributing to the electron EDM are depicted in Fig.~\ref{Feynman diagram}. Calculating the Feynman diagrams, $d_e$ at the one-loop level can be written as
\begin{eqnarray}
	&&d_e^{(1)}=\frac{-2}{em_e}\Im\Big\{x_e[-I_3(x_{F_j},x_{S_i})+I_4(x_{F_j},x_{S_i})][(C_{\bar l_e S_i F_j}^LC_{\bar F_j S_i l_e}^R)+(C_{\bar l_e S_i F_j}^RC_{\bar F_j S_i l_e}^L)^*]\nonumber\\
	&&\qquad\quad+\sqrt{x_{e}x_{F_j}}[-2I_1(x_{F_j},x_{S_i})+2I_3(x_{F_j},x_{S_i})]C_{\bar l_e S_i F_j}^RC_{\bar F_j S_i l_e}^R\Big\},\nonumber\\
	&&d_e^{(2)}=\frac{-2}{em_e}\Im\Big\{x_e[-I_1(x_{F_j},x_{S_i})+2I_3(x_{F_j},x_{S_i})-I_4(x_{F_j},x_{S_i})][(C_{\bar l_e S_i F_j}^RC_{\bar F_j S_i l_e}^L)\nonumber\\
	&&\qquad\quad+(C_{\bar l_e S_i F_j}^LC_{\bar F_j S_i l_e}^R)^*]+\sqrt{x_{e}x_{F_j}}[2I_1(x_{F_j},x_{S_i})-2I_2(x_{F_j},x_{S_i})-2I_3(x_{F_j},x_{S_i})]\nonumber\\
	&&\qquad\quad\times C_{\bar l_e S_i F_j}^RC_{\bar F_j S_i l_e}^R\Big\},\nonumber\\
	&&d_e=d_e^{(1)}+d_e^{(2)},
\end{eqnarray}
where $x_i=m_i^2/m_W^2$, $C_{abc}^{L,R}$ denotes the constant parts of the interactional vertex about $abc$, which can be got through SARAH, the interacting particles are denoted by $a$, $b$, $c$. And the specific expressions for the functions
$I_{1,2,3,4}$ is given by Refs.~\cite{Zhang:2013hva,Zhang:2013jva}. For completeness, we will provide the specific forms of $C_{abc}^{L,R}$ and $I_{1,2,3,4}$ in Appendix~\ref{constants}.

	\section{Numerical analyses\label{sec4}}
	
In this section, we provide the numerical results of EDMs $d_n$, $d_b$, $d_c$, $d_e$ and $d_{Hg}$ in the TNMSSM. For SM parameters, we take the W boson mass $m_W=80.337{\;\rm GeV}$, the Z boson mass $m_Z=91.1876{\;\rm GeV}$, the $u$ quark mass $m_u=2.2{\;\rm MeV}$, the $d$ quark mass$m_d=4.7{\;\rm MeV}$, the $b$ quark mass $m_b=4.18{\;\rm GeV}$, the $c$ quark mass $m_c=1.27{\;\rm GeV}$, $\alpha_{em}(m_Z)=1/128.9$ for the coupling of the electromagnetic interaction, $\alpha_s(m_Z)=0.118$ for the coupling of the strong
interaction \cite{17,CMS.PLB,ATLAS.PLB,PC,MV,MC,FW,MC1,ME}. According to the analysis in Ref.~\cite{Yan:2020ocy, Agashe:2011ia,Yang:2024uoq}, we take $\tan\beta'=1.3$, $v_T^2+v_{\bar T}^2=1 {GeV}^2$. According to the analysis in Ref.~\cite{gluino}, we take the mass of gluino larger than 2 TeV. For the squark sector, we take $m_{\tilde q}=m_{\tilde u}=m_{\tilde d}=diag(M_Q,M_Q,M_Q){\;\rm TeV}$ and $T_{u,d}=Y_{u,d}\;diag(A_Q,A_Q,A_Q) {\;\rm TeV}$ for simplicity. The observed Higgs signal limits that $M_Q>1.5\;{\rm TeV}$~\cite{Un:2016hji}. Considering the experimental constraints described in Eq.~(\ref{EDMlimits}) and Eq.~(\ref{EDMbc}), we adopt the following parameters to carry out the numerical calculation
\begin{eqnarray}\nonumber
	&& \tan\beta=5,\;\tan\beta'=1.3,\;\lambda_{T}=-0.22,\;\kappa=0.8,\;\chi_d=\chi_t=0.1,\nonumber\\
	&&T_{\lambda}=T_{\lambda_T}=-0.1~{\rm TeV},\;|M_1|=0.5~{\rm TeV},\;|M_2|=0.6~{\rm TeV},\nonumber\\
	&&T_{\chi_t}=-0.8~{\rm TeV},\;T_{\chi_d}=-0.5~{\rm TeV}.\
	\label{parameter}
\end{eqnarray}

\begin{figure}
	\setlength{\unitlength}{1mm}
	\centering
	\includegraphics[width=2.0in]{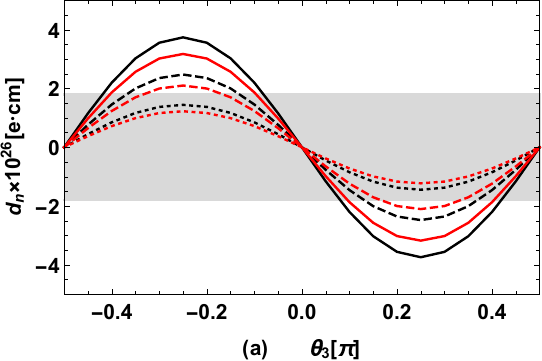}%
	\vspace{0.3cm}
	\includegraphics[width=2.0in]{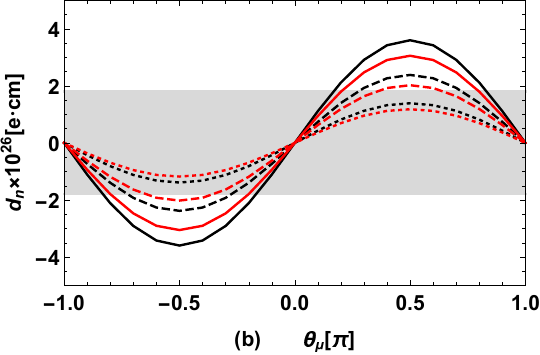}
	\vspace{0cm}
	\par
	\hspace{-0.in}
	\includegraphics[width=2.0in]{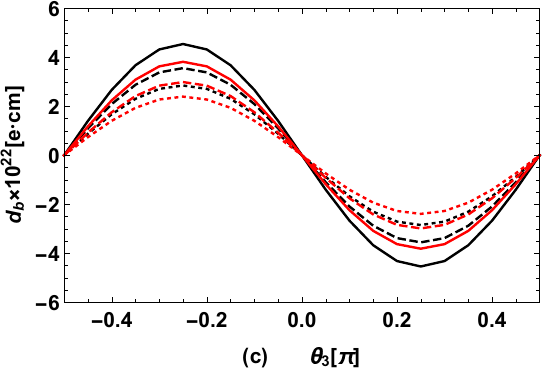}%
	\vspace{0.3cm}
	\includegraphics[width=2.0in]{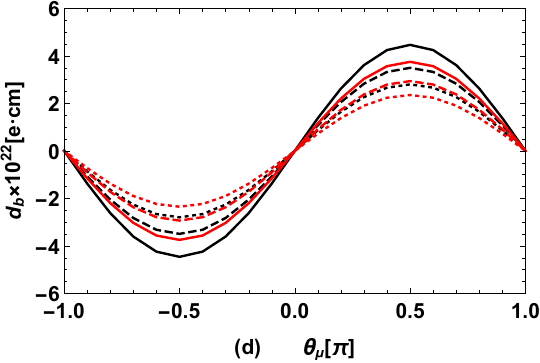}
	\vspace{0cm}
	\par
	\hspace{-0.in}
	\includegraphics[width=2.0in]{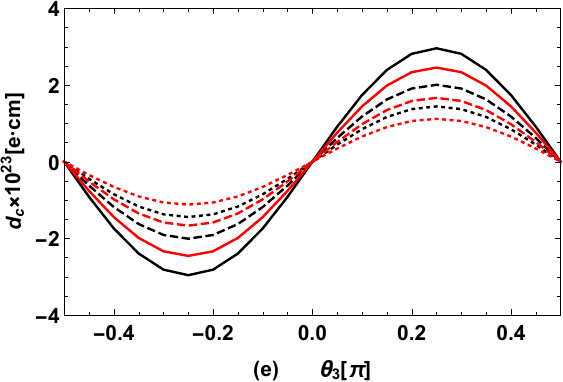}%
	\vspace{0.3cm}
	\includegraphics[width=2.0in]{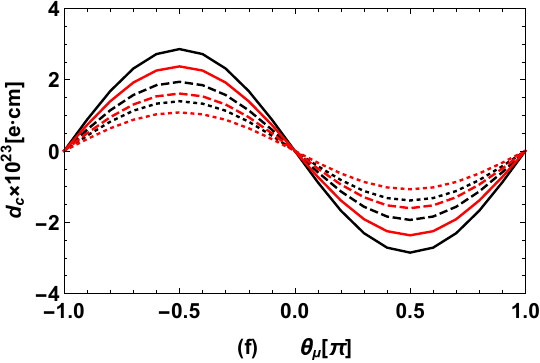}
	\vspace{0cm}
	\par
	\hspace{-0.in}
	\includegraphics[width=2.0in]{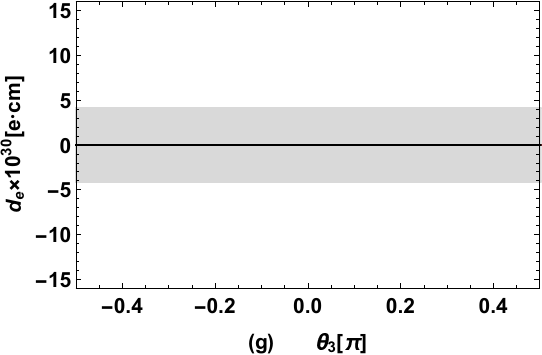}%
	\vspace{0.3cm}
	\includegraphics[width=2.0in]{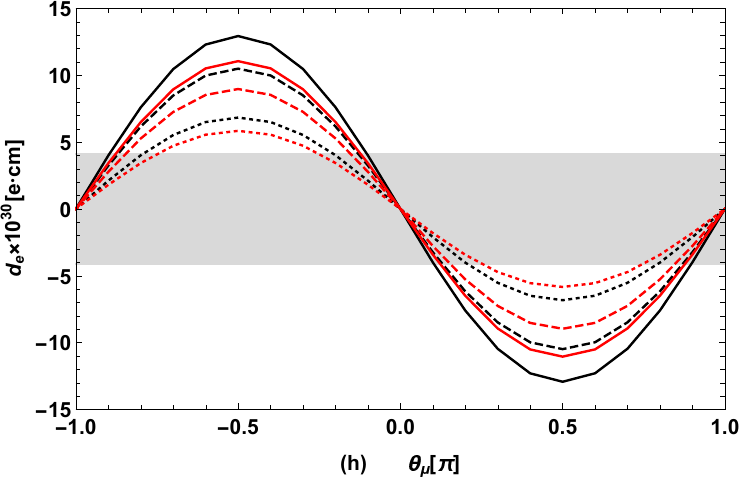}
	\vspace{0cm}
	\par
	\hspace{-0.in}
	\includegraphics[width=2.0in]{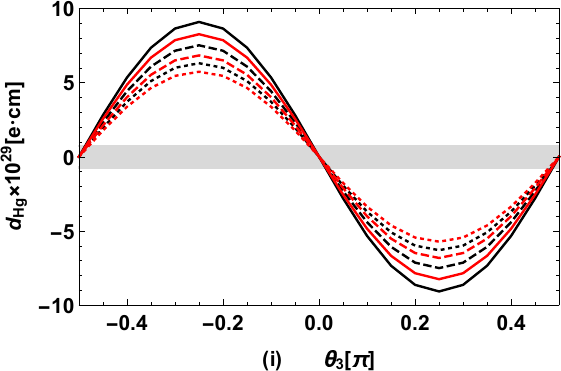}%
	\vspace{0.3cm}
	\includegraphics[width=2.0in]{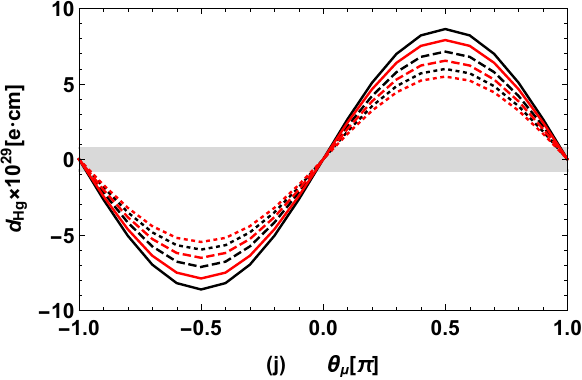}
	\vspace{0cm}
	
	\caption[]{Take the values from Eq.~(\ref{parameter}), $d_n$, $d_b$, $d_c$, $d_e$,$d_{Hg}$ versus $\theta_3$, $\theta_\mu$ are plotted, where the solid, dashed and dotted lines denote the results for $M_3=4.00,\;4.25,\;4.50\;{\rm TeV}$ respectively for (a), (c), (e), (g), (i) and the results for $\mu=1.2,\;1,\;0.8\;{\rm TeV}$ respectively for (b), (d), (f), (h), (j). Similarly, the results predicted in the MSSM with the same parameter space are denoted by the red lines.}
	\label{theta3mu}
\end{figure}

	Then we calculate the contributions from one loop diagrams and the Weinberg operators of neutron ($d_n$), $b$ quark ($d_b$), $c$ quark ($d_c$), electron ($d_e$) and mercury ($d_{Hg}$). To explore the effects of $\theta_3$ (the phase angle of gluino mass $M_3$) and $\mu$ term makes the dominant contributions to the EWB ~\cite{King:2015oxa}, and the corresponding CPV phase $\theta_\mu$ is requested to be large.  We plot $d_n$ versus $\theta_3$ in Fig.~\ref{theta3mu} (a) and $\theta_ \mu$ in Fig.~\ref{theta3mu} (b), where the shaded area represent the experimental range of $d_n$, the black solid, black dashed, black dotted lines in in Fig.~\ref{theta3mu} (a) denotes the results for $M_3=4.00,\;4.25,\;4.50\;{\rm TeV}$ respectively, the black solid, black dashed, black dotted lines in in Fig.~\ref{theta3mu} (b) denotes the results for $\mu=1.2,\;1,\;0.8\;{\rm TeV}$ respectively. Similarly, we plot $d_b$ versus $\theta_3$ in Fig.~\ref{theta3mu} (c), $d_b$ versus $\theta_ \mu$ in Fig.~\ref{theta3mu} (d), $d_c$ versus $\theta_3$ in Fig.~\ref{theta3mu} (e), $d_c$ versus $\theta_ \mu$ in Fig.~\ref{theta3mu} (f), $d_e$ versus $\theta_3$ in Fig.~\ref{theta3mu} (g), $d_e$ versus $\theta_ \mu$ in Fig.~\ref{theta3mu} (h), $d_{Hg}$ versus $\theta_3$ in Fig.~\ref{theta3mu} (i), $d_{Hg}$ versus $\theta_ \mu$ in Fig.~\ref{theta3mu} (j). The red lines in Fig.~\ref{theta3mu} denote the results predicted in the MSSM with the same parameter space.

From the picture, it is obvious that the results for $d_n$, $d_b$, $d_c$, $d_e$, $d_{Hg}$ predicted in the TNMSSM are larger than the ones predicted in the MSSM, because the additional neutralinos, charginos, the redefinitions of squark mass matrices and the $\mu$ term can make contributions to these EDMs. In addition, the effects of $\theta_3$, $\theta_\mu$ on  $d_n$, $d_b$, $d_c$, $d_e$, $d_{Hg}$ can be amplified for larger $M_3$ and larger $\mu$ respectively, it indicates these EDMs increase with increasing $M_3$, $\mu$ for fixed $\theta_3$ and $\theta_\mu$, the fact can be seed explicitly in Fig.~\ref{theta3mu}. Due to the strict upper bound on $d_n$ experimentally, $\theta_3$, $\theta_\mu$ are limited to be very small for large $M_3$ and $\mu$ respectively, while they are not constrained by the upper bounds on $d_b$, $d_c$, $d_e$, $d_{Hg}$ in the same parameter space as shown in Fig.~\ref{theta3mu} (c), (d), (e), (f), (g), (h), (i), (j).

\begin{figure}
	\setlength{\unitlength}{1mm}
	\centering
	\includegraphics[width=2.0in]{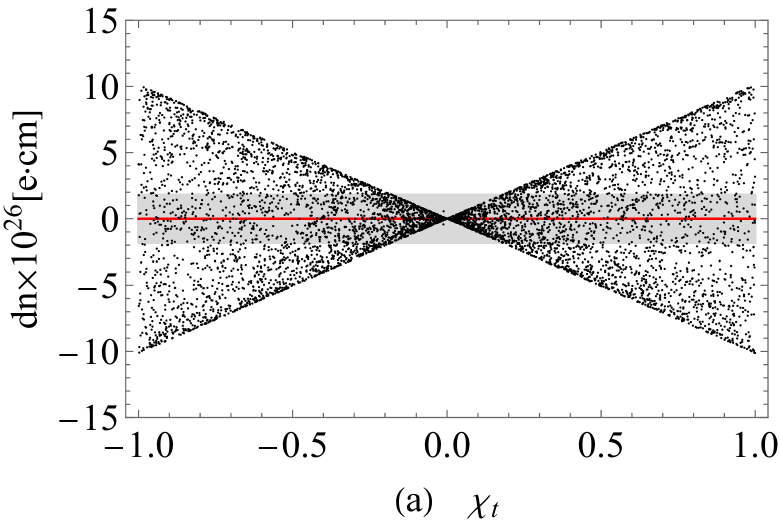}%
	\vspace{0.5cm}
	\includegraphics[width=2.0in]{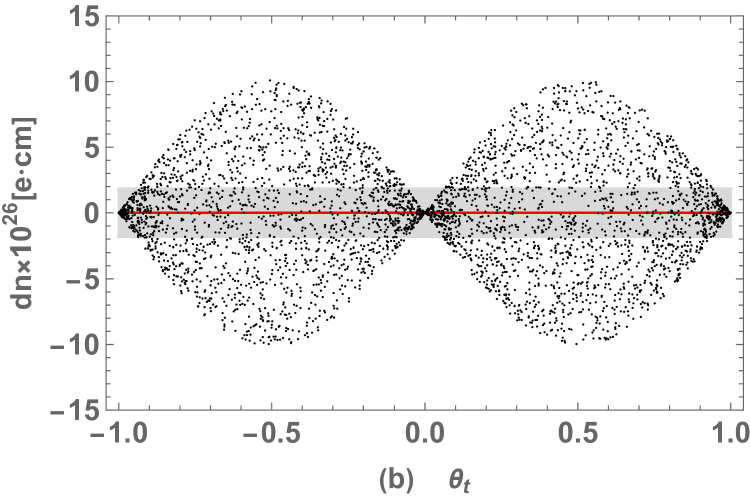}
	\vspace{0cm}
	\par
	\hspace{-0.in}
	\includegraphics[width=2.0in]{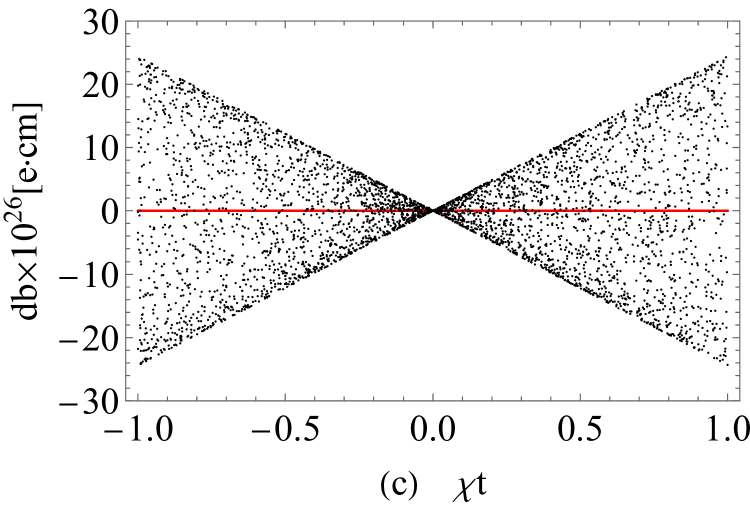}%
	\vspace{0.5cm}
	\includegraphics[width=2.0in]{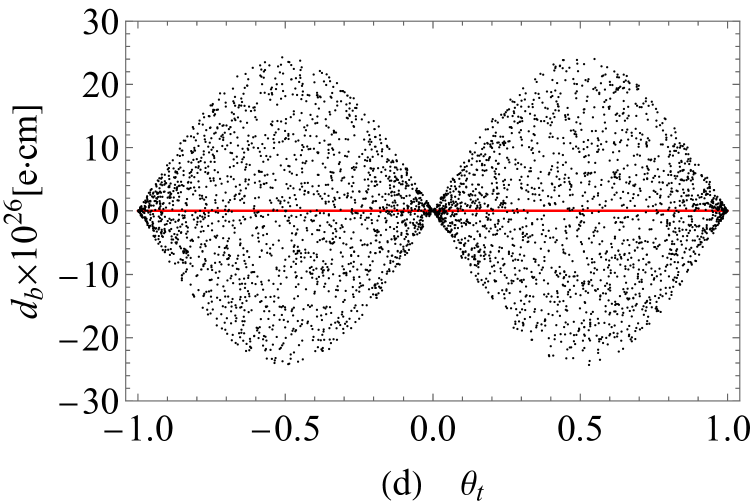}
	\vspace{0cm}
	\par
	\hspace{-0.in}
	\includegraphics[width=2.0in]{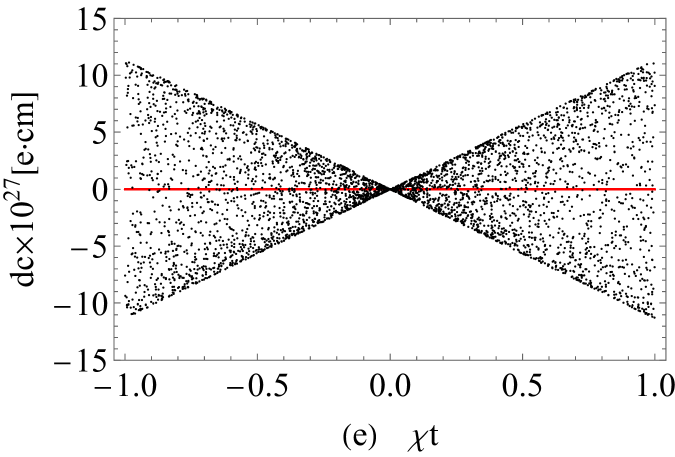}%
	\vspace{0.5cm}
	\includegraphics[width=2.0in]{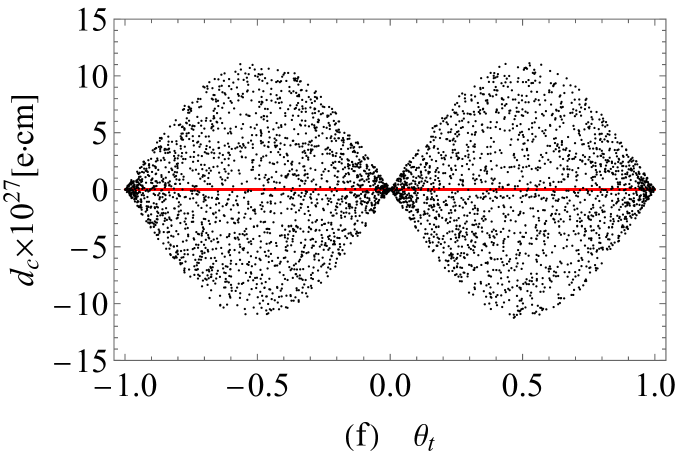}
	\vspace{0cm}
	\par
	\hspace{-0.in}
	\includegraphics[width=2.0in]{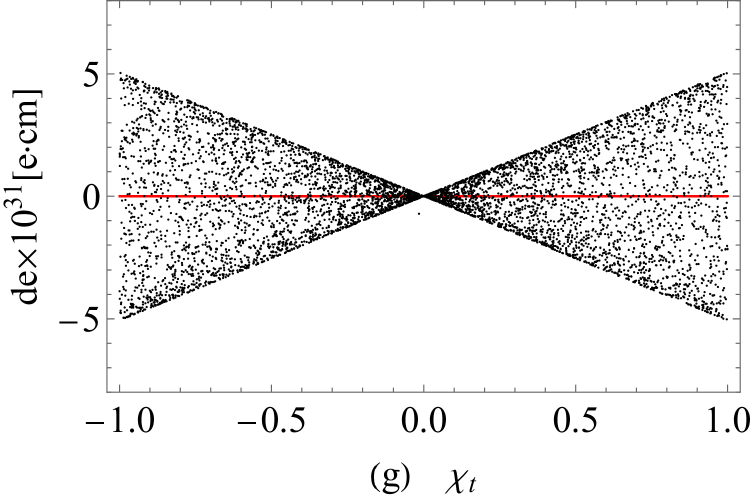}%
	\vspace{0.5cm}
	\includegraphics[width=2.0in]{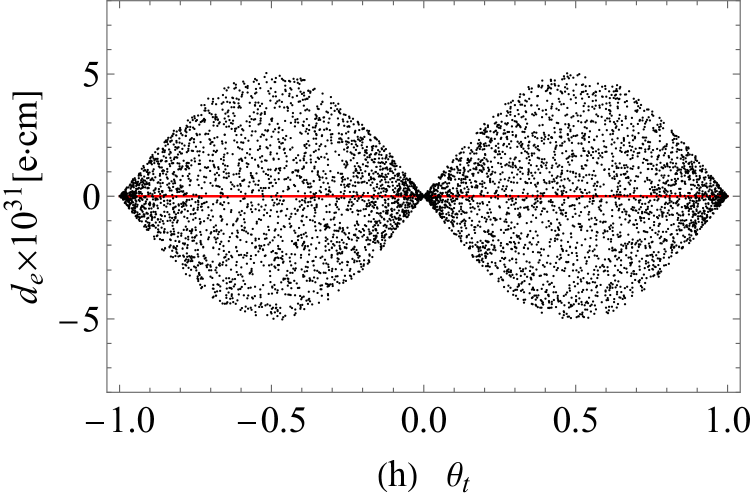}
	\vspace{0cm}
	\par
	\hspace{-0.in}
	\includegraphics[width=2.0in]{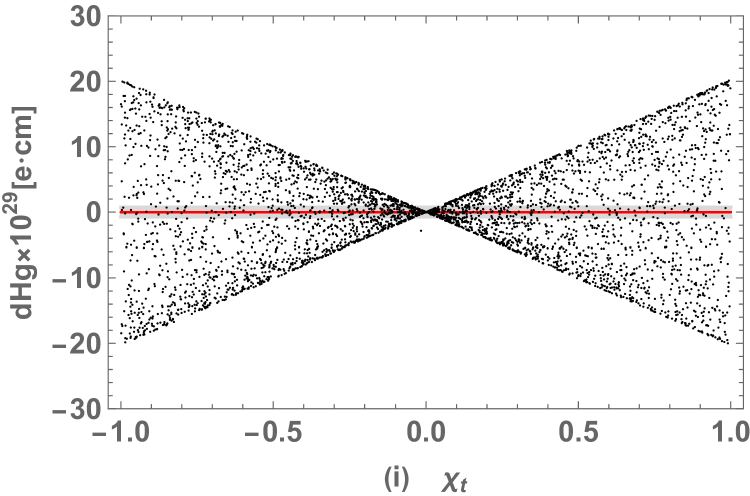}%
	\vspace{0.5cm}
	\includegraphics[width=2.0in]{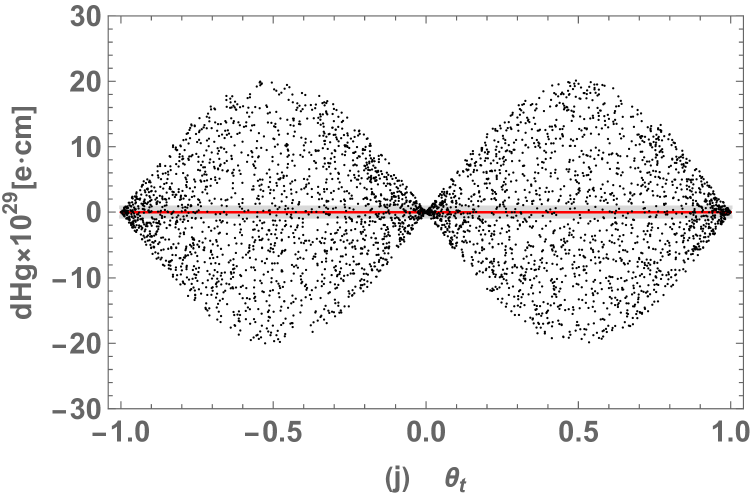}
	\vspace{0cm}
	\caption[]{Scanning the parameter space shown in Eq.~(\ref{R3}), $d_n$, $d_b$, $d_c$, $d_e$ and $d_{Hg}$ versus $\chi_t$, $\theta_t$ are plotted, where the red lines denote the corresponding MSSM predictions in the same parameter space and the shaded areas represent the experimental upper bound on $d_n$.}
	\label{T}
\end{figure}

\begin{figure}
	\setlength{\unitlength}{1mm}
	\centering
	\includegraphics[width=2.0in]{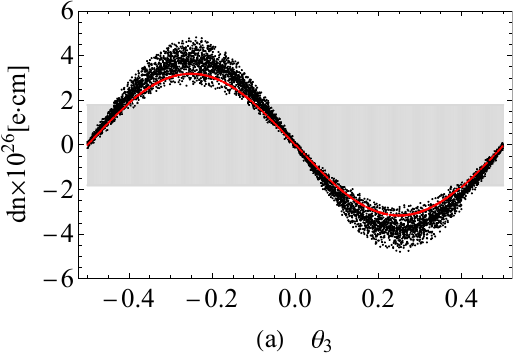}%
	\vspace{0.5cm}
	\includegraphics[width=2.0in]{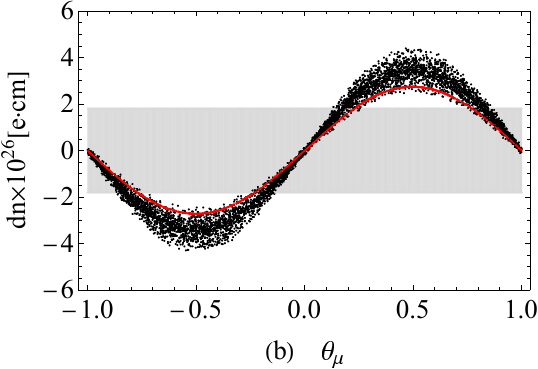}
	\vspace{0cm}
	\par
	\hspace{-0.in}
	\includegraphics[width=2.0in]{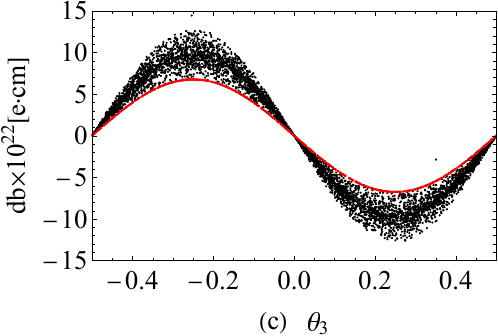}%
	\vspace{0.5cm}
	\includegraphics[width=2.0in]{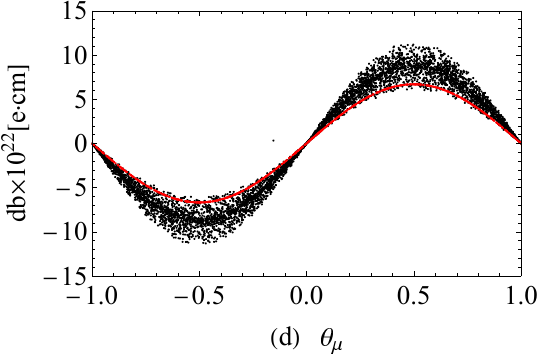}
	\vspace{0cm}
	\par
	\hspace{-0.in}
	\includegraphics[width=2.0in]{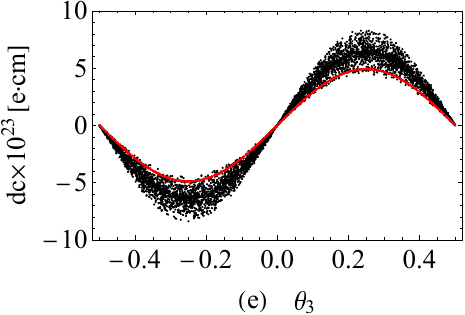}%
	\vspace{0.5cm}
	\includegraphics[width=2.0in]{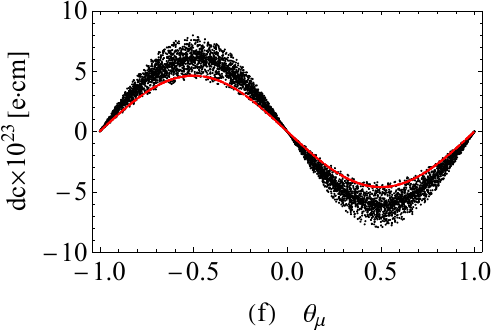}
	\par
	\hspace{-0.in}
    \includegraphics[width=2.0in]{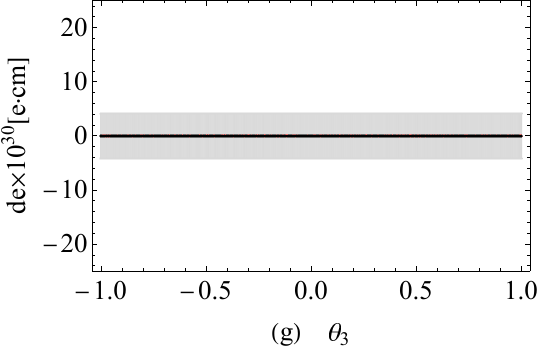}%
	\vspace{0.5cm}
	\includegraphics[width=2.0in]{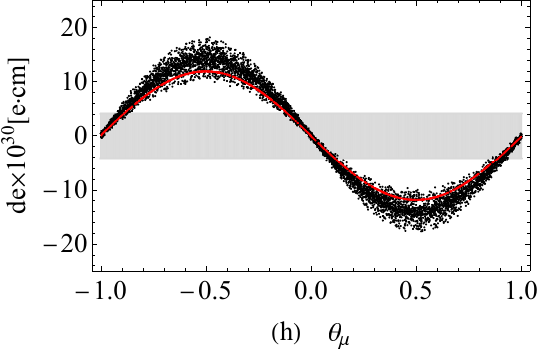}
	\par
	\hspace{-0.in}
	\includegraphics[width=2.0in]{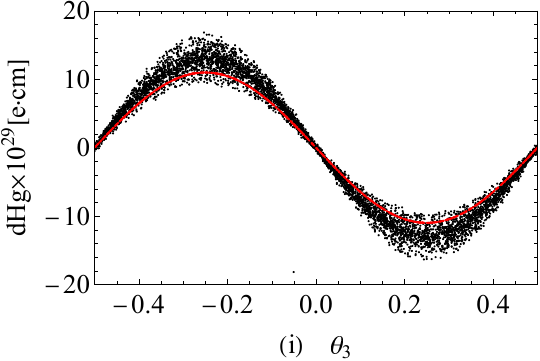}%
	\vspace{0.5cm}
	\includegraphics[width=2.0in]{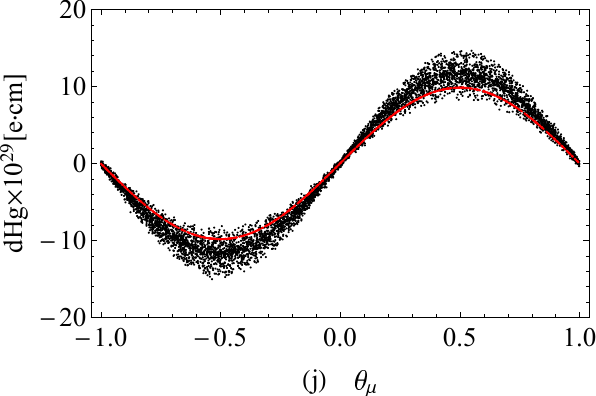}
	\vspace{0cm}
	\caption[]{Scanning the parameter space shown in Eq.~(\ref{R3}), $d_n$, $d_b$, $d_c$, $d_e$ and $d_{Hg}$ versus $\theta_3$, $\theta_\mu$ are plotted, where the red lines denote the corresponding MSSM predictions in the same parameter space and the shaded areas represent the experimental range of $d_n$ and $d_e$}.
	\label{TN}
\end{figure}
Comparing with the MSSM, there are two new CPV sources $\chi_d$ and $\chi_t$ in the TNMSSM, which can also make contributions to the considered EDMs $d_n$, $d_b$, $d_c$, $d_e$, $d_{Hg}$. To see explicitly their effects on $d_n$, $d_b$, $d_c$, $d_e$, $d_{Hg}$, we assume all contributions to these EDMs come from $\theta_d$, $\theta_t$ and scan the the following parameter space
\begin{eqnarray}
	&&\chi_d=(-1,\;1),\chi_t=(-1,\;1),\;\theta_d=(-\pi,\;\pi),\;\theta_t=(-\pi,\;\pi).\label{R3}
\end{eqnarray}
Taking the parameters in Eq.~(\ref{parameter}) as inputs, we plot $d_n$ versus $\chi_t$, $\theta_t$ in Fig.~\ref{T} (a), (b) respectively, where the shaded area represent the experimental range of $d_n$, the red lines denote the results predicted in the MSSM with the same parameter space (which is $0$ in this case because there is no such CPV sources in the MSSM). Similarly, we plot $d_b$, $d_c$, $d_e$ and $d_{Hg}$ versus  $\chi_t$, $\theta_t$ in Fig.~\ref{T} (c, d), (e, f), (g, h), (i, j) respectively. There are no shaded area in Fig.~\ref{T} (c, d), (e, f), (g, h), (i, j) because the upper bounds on $d_b$, $d_c$, $d_e$ and $d_{Hg}$ are much larger than the results shown in the picture.

From Eq.~(\ref{9}) to Eq.~(\ref{14}), it can be noted that the effects of $\chi_t$ are enhanced by $\tan\beta$ compared to the ones of $\chi_d$, and dominant contributions to these EDMs come from the up-type squarks which related directly to $\chi_t$. As a result, $\chi_t$ affects the numerical results of $d_n$, $d_b$, $d_c$, $d_e$, $d_{Hg}$ much more acutely than the ones of $\chi_d$, the fact can be seen explicitly in Fig.~\ref{T} because the predicted $d_n$, $d_b$, $d_c$, $d_e$, $d_{Hg}$ approach to $0$ when $\theta_t=0$. Fig.~\ref{T} (a, b) show that the new CPV source $\chi_t$ also suffer the constraints from the experimental upper bound on $d_n$, and the constraints on $\theta_t$ can be relaxed when $|\chi_t|\lesssim0.2$. In addition, the theoretical predictions on $d_b$, $d_c$ can reach $10^{-26}$ e$\cdot$cm, $10^{-27}$ e$\cdot$cm, $d_e$, $d_{Hg}$ can reach $10^{-31}$ e$\cdot$cm, $10^{-29}$ e$\cdot$cm respectively when the contributions come from the new CPV effects in the TNMSSM, which has the potential to be verified experimentally in future.

To see the combining effects of the new CPV sources $\chi_d$, $\chi_t$ in the TNMSSM and the traditional CPV sources $M_3$, $\mu$ in SUSY models on $d_n$, $d_b$, $d_c$, $d_e$, $d_{Hg}$, we set $M_3=5\;{\rm TeV}$, $\mu=0.6$ and scan the parameter space in Eq.~(\ref{R3}). Taking $\theta_\mu=0$, we plot $d_n$ versus $\theta_3$ in Fig.~\ref{TN} (a), where the red line denotes the corresponding MSSM prediction in the same parameter space, the shaded area represents the experimental upper bound on $d_n$. Similarly, $d_n$ versus $\theta_\mu$ for $\theta_3=0$ are plotted in Fig.~\ref{TN} (b), and $d_b$ versus $\theta_3$, $d_b$ versus $\theta_\mu$, $d_c$ versus $\theta_3$, $d_c$ versus $\theta_\mu$, $d_e$ versus $\theta_3$, $d_e$ versus $\theta_\mu$, $d_{Hg}$ versus $\theta_3$, $d_{Hg}$ versus $\theta_\mu$ are plotted in Fig.~\ref{TN} (c), (d), (e), (f), (g), (h), (i), (j) respectively, the shaded area represents the experimental upper bound on $d_e$. The picture shows explicitly that the new CPV sources in the TNMSSM can make important contributions to the EDMs $d_n$, $d_b$, $d_c$, $d_e$, $d_{Hg}$, and these EDMs predicted in the TNMSSM are much larger than the ones predicted in the MSSM shown by the red lines in Fig.~\ref{TN}. In addition, the effects of the new CPV sources $\chi_d$, $\chi_t$ are related closely to $M_3$ and $\mu$, because they affect the numerical results slightly when $\theta_3$ or $\theta_\mu$ approaches to $0$ as shown in Fig.~\ref{TN}. By comparing the (g, h), (i, j) with (a, b) in Fig.~\ref{theta3mu} and Fig.~\ref{TN}, it can be observed that the phase $\theta_\mu$ of the coupling $\mu$ is quite constrained by the experimental limits on the electron EDM and mercury EDM. And $d_e$, $d_{Hg}$ also have stronger restrictions on $\theta_\mu$ than that on $d_n$.

	\section{Summary\label{sec5}}
	
In this work, we study and analyze the EDMs of neutron, electron, mercury and heavy quarks $b$, $c$ in the TNMSSM. The redefined $\mu$ term and squark mass matrices, the new introduced CPV sources $\chi_d$, $\chi_t$ in the TNMSSM can make contributions to these EDMs. We calculate the loop-induced contributions through the effective theory, and all contributions are presented as the coefficients of effective operators. Considering the latest upper bounds on the EDMs $d_n$, $d_e$, $d_{Hg}$, $d_b$, $d_c$ and focusing on the TNMSSM-specialized CPV sources, we carry out the numerical analysis properly. It is found that the theoretical predictions on $d_n$, $d_e$, $d_{Hg}$, $d_b$, $d_c$ in the TNMSSM are much different from the ones in the MSSM due to the presence of new particles and new CPV sources in the TNMSSM, where the new CPV source $\chi_t$ in the TNMSSM can make significant contributions to these EDMs. In addition, the parameter space of the TNMSSM can be limited by the experimental upper bound on $d_n$, $d_e$ and $d_{Hg}$, the predicted $d_b$, $d_c$ can reach $10^{-22}$ e$\cdot$cm, $10^{-23}$ e$\cdot$cm respectively which have the potential to be verified experimentally in future\cite{Yan:2020ocy, Cheung:2011wn,RuizVidal:2022yug}.

	\begin{acknowledgments}
		
The work has been supported by the National Natural Science Foundation of China (NNSFC) with Grants No. 12075074, No. 12235008, Hebei Natural Science Foundation with Grant No. A2022201017, No. A2023201041, Natural Science Foundation of Guangxi Autonomous Region with Grant No. 2022GXNSFDA035068, the youth top-notch talent support program of the Hebei Province.		
		
	\end{acknowledgments}

	\appendix
\section{New definitions of some mass matrixes in the TNMSSM. \label{mass matrix}}

In the TNMSSM, the dominant contributions to $d_n$, $d_b$, $d_c$ come from squarks, sgluon, neutralinos and charginos. To see the new contributions in the model to these EDMS clearly, we present the new definitions of the relevant mass matrixes. On the basis $\left(\tilde{d}^0_{L,{{\alpha_1}}}, \tilde{d}^0_{R,{{\alpha_2}}}\right), \left(\tilde{d}^{0,*}_{L,{{\beta_1}}}, \tilde{d}^{0,*}_{R,{{\beta_2}}}\right)$, the definition of the squared mass matrix for down type squark is given by ~\cite{SAR1}
\begin{equation}
	M^2_{\tilde{D}} = \left(
	\begin{array}{cc}
		m_{\tilde{d}_L^0\tilde{d}_L^{0,*}} &m^\dagger_{\tilde{d}_R^0\tilde{d}_L^{0,*}}\\
		m_{\tilde{d}_L^0\tilde{d}_R^{0,*}} &m_{\tilde{d}_R^0\tilde{d}_R^{0,*}}\end{array}
	\right),
	\label{9}
\end{equation}
where
\begin{eqnarray}
	&&m_{\tilde{d}_L^0\tilde{d}_L^{0,*}} = -\frac{1}{24}
	\Big( (3 g_{2}^{2}  + g_{1}^{2}) (2 v_{\bar{T}}^{2}  -2 v_{T}^{2}  - v_{u}^{2}  + v_{d}^{2})\Big)+\frac{1}{2}(2m_{{q}}^2  +{ v_{d}^{2} Y_dY_{d}^{\dagger}}),\nonumber\\
	&&m_{\tilde{d}_L^0\tilde{d}_R^{0,*}} = \frac{1}{2}  \Big(\sqrt{2}v_d T_d +Y_d(2v_d v_T{\chi}_{d}^* - v_s v_u  {\lambda}^*) \Big),\nonumber\\
	&&m_{\tilde{d}_R^0\tilde{d}_R^{0,*}} = \frac{1}{12}
	g_{1}^{2} (-2 v_{\bar{T}}^{2} +2 v_{T}^{2}  - v_{d}^{2}  + v_{u}^{2})+\frac{1}{2}(2m_{{d}}^2  +{ v_{d}^{2}Y_{d}  Y_d}^{\dagger}).
	\label{10}
\end{eqnarray}

On the basis $\left(\tilde{u}^0_{L,{{\alpha_1}}}, \tilde{u}^0_{R,{{\alpha_2}}}\right), \left(\tilde{u}^{0,*}_{L,{{\beta_1}}}, \tilde{u}^{0,*}_{R,{{\beta_2}}}\right)$, the squared mass matrix of up type squarks is~\cite{SAR2}
\begin{equation}
	M^2_{\tilde{U}} = \left(
	\begin{array}{cc}
		m_{\tilde{u}_L^0\tilde{u}_L^{0,*}} &m^\dagger_{\tilde{u}_R^0\tilde{u}_L^{0,*}}\\
		m_{\tilde{u}_L^0\tilde{u}_R^{0,*}} &m_{\tilde{u}_R^0\tilde{u}_R^{0,*}}\end{array}
	\right),
	\label{11}
\end{equation}
where
\begin{eqnarray}
	&&m_{\tilde{u}_L^0\tilde{u}_L^{0,*}} = -\frac{1}{24}
	\Big( (-3 g_{2}^{2}  + g_{1}^{2}) (2 v_{\bar{T}}^{2}  -2 v_{T}^{2}  - v_{u}^{2}  + v_{d}^{2})\Big)+\frac{1}{2}(2m_{{q}}^2  +{ v_{u}^{2} Y_uY_{u}^{\dagger}}),\nonumber\\
	&&m_{\tilde{u}_L^0\tilde{u}_R^{0,*}} =  \frac{1}{2}  \Big(\sqrt{2}v_u T_u +Y_u(2 v_{\bar{T}}v_u{\chi}_{t}^* - v_d v_s  {\lambda}^*) \Big),\nonumber\\
	&&m_{\tilde{u}_R^0\tilde{u}_R^{0,*}} =\frac{1}{6}
	g_{1}^{2} (2 v_{\bar{T}}^{2} -2 v_{T}^{2}  - v_{u}^{2}  + v_{d}^{2})+\frac{1}{2}(2m_{{u}}^2  +{ v_{u}^{2}Y_{u}  Y_u}^{\dagger}).
	\label{12}
\end{eqnarray}

On the basis $(\tilde\lambda_B, \tilde{W}^0, \tilde{H_d^0}, \tilde{H_u^0}, \tilde{S}, \tilde{T^0}, \tilde{\bar T^0})$, the mass matrix for neutralinos is~\cite{SAR3}
\begin{eqnarray}
	&&m_{\chi^0}^2=
	\left(\begin{array}{ccccccc}M_1 &0 &-\frac{1}{2}g_1v_d &\frac{1}{2}g_1v_u & 0 &g_1v_T &-g_1v_{\bar{T}}\\
		0 &M_2 &\frac{1}{2}g_2v_d &-\frac{1}{2}g_2v_u &0 &-g_2v_T &g_2v_{\bar{T}}\\
		-\frac{1}{2}g_1v_d &\frac{1}{2}g_2v_d &\sqrt{2}v_T{\chi}_d &-\frac{1}{\sqrt{2}}v_s \lambda &-\frac{1}{\sqrt{2}}v_u \lambda &\sqrt{2}v_d{\chi}_d &0\\
		\frac{1}{2}g_1v_u &-\frac{1}{2}g_2v_u &-\frac{1}{\sqrt{2}}v_s  \lambda &\sqrt{2}v_{\bar{T}}{\chi}_t &-\frac{1}{\sqrt{2}}v_d \lambda &0 &\sqrt{2}v_u{\chi}_t\\
		0 &0 &-\frac{1}{\sqrt{2}}v_u \lambda &-\frac{1}{\sqrt{2}}v_d \lambda &\sqrt{2}v_s\kappa &-\frac{1}{\sqrt{2}}\lambda_Tv_{\bar{T}} &-\frac{1}{\sqrt{2}}\lambda_Tv_T\\
		g_1v_T &-g_2v_T &\sqrt{2}v_d{\chi}_d &0 &-\frac{1}{\sqrt{2}}\lambda_Tv_{\bar{T}} &0 &-\frac{1}{\sqrt{2}}\lambda_Tv_s\\
		-g_1v_{\bar{T}} &g_2v_{\bar{T}} &0 &\sqrt{2}v_u{\chi}_t &-\frac{1}{\sqrt{2}}\lambda_Tv_T &-\frac{1}{\sqrt{2}}\lambda_Tv_s &0
	\end{array}\right).
	\label{13}
\end{eqnarray}

On the basis $ \left(\tilde{W}^-, \tilde{H}_d^-,\tilde{T}^-\right), \left(\tilde{W}^+, \tilde{H}_u^+,\tilde{T}^+\right)$, the mass matrix for charginos is~\cite{SAR4,SAR5}
\begin{equation}
	m_{\tilde{\chi}^-} = \left(
	\begin{array}{ccc}
		M_2 &\frac{1}{\sqrt{2}} g_2 v_u & g_2 v_T\\
		\frac{1}{\sqrt{2}} g_2 v_d  &\frac{1}{\sqrt{2}} v_s \lambda & -v_d {\chi}_d\\
		g_2 v_{\bar{T}} & -v_u {\chi}_t &\frac{1}{\sqrt{2}} \lambda_T v_s\end{array}
	\right).
	\label{14}
\end{equation}
In the basis $(\phi_d,\phi_u,\phi_s,\phi_T,\phi_{\bar{T}})$, the definition of mass squared matrix for neutral Higgs is given by
\begin{eqnarray}
	m_h^2=\begin{pmatrix}
		m_{\phi_d \phi_d}&m_{\phi_u \phi_d}&m_{\phi_s \phi_d}&m_{\phi_T \phi_d}&m_{\phi_{\bar{T}} \phi_d}\\
		m_{\phi_d \phi_u}&m_{\phi_u \phi_u}&m_{\phi_s \phi_u}&m_{\phi_T \phi_u}&m_{\phi_{\bar{T}} \phi_u}\\
		m_{\phi_d \phi_s}&m_{\phi_u \phi_s}&m_{\phi_s \phi_s}&m_{\phi_T \phi_s}&m_{\phi_{\bar{T}} \phi_s}\\
		m_{\phi_d \phi_T}&m_{\phi_u \phi_T}&m_{\phi_s \phi_T}&m_{\phi_T \phi_T}&m_{\phi_{\bar{T}} \phi_T}\\
		m_{\phi_d \phi_{\bar{T}}}&m_{\phi_u \phi_{\bar{T}}}&m_{\phi_s \phi_{\bar{T}}}&m_{\phi_T \phi_{\bar{T}}}&m_{\phi_{\bar{T}} \phi_{\bar{T}}}\\
	\end{pmatrix}
\end{eqnarray}
where
\begin{align}
	m_{\phi_d \phi_d}&=m_{H_d}^2+\frac{1}{8}(g_1^2+g_2^2)(2v_{\bar{T}}^2-2v_T^2+3v_d^2-v_u^2)\nonumber\\
	&+\sqrt{2}v_T\Re(T_{\chi_d})+\frac{|\lambda|^2}{2}(v_s^2+v_u^2)-v_s v_{\bar{T}} \Re(\chi_d \Lambda_T^*)+(3v_d^2+2v_T^2)|\chi_d|^2,\nonumber\\	
	m_{\phi_d \phi_u}&=-\frac{1}{4}(g_1^2+g_2^2)v_d v_u-\frac{1}{\sqrt{2}}v_s \Re(T_\lambda)
	+\frac{1}{2}v_T v_{\bar{T}} \Re(\lambda\Lambda_T^*)+v_d v_u |\lambda|^2\nonumber\\
	&-v_s v_{\bar{T}}\Re(\chi_t \lambda^*)-v_s v_T\Re(\chi_d \lambda^*)-\frac{1}{2}v_s^2\Re(\kappa\lambda^*),\nonumber\\
	m_{\phi_u \phi_u}&=m_{H_u}^2-\frac{1}{8}(g_1^2+g_2^2)(2v_{\bar{T}}^2-2v_T^2-3v_u^2+v_d^2)\nonumber\\
	&+\sqrt{2}v_{\bar{T}}\Re(T_{\chi_t})+\frac{1}{2}(v_d^2+v_s^2)|\lambda|^2-v_s v_T \Re(\chi_t \Lambda_T^*)+(2v_{\bar{T}}^2+3v_u^2)|\chi_t|^2,\nonumber\\
	m_{\phi_d \phi_s}&=v_d v_s|\lambda|^2-\frac{1}{\sqrt{2}}v_u \Re(T_\lambda)-v_s v_u \Re(\kappa\lambda^*)
	-v_T v_u \Re(\lambda\chi_d^*)-v_{\bar{T}}v_u \Re(\chi_t \lambda^*)-v_d v_{\bar{T}} \Re(\Lambda_T \chi_d^*),\nonumber\\
	m_{\phi_u \phi_s}&=v_s v_u|\lambda|^2-\frac{1}{\sqrt{2}}v_d\Re(T_\lambda)-v_d v_s\Re(\lambda\kappa^*)
	-v_T v_d\Re(\lambda\chi_d^*)-v_d v_{\bar{T}}\Re(\lambda\chi_t^*)-v_T v_u\Re(\chi_t \Lambda_T^*),\nonumber\\
	m_{\phi_s \phi_s}&=m_S^2+\frac{|\Lambda_T|^2}{2}(v_T^2+v_{\bar{T}}^2)-v_T
	v_{\bar{T}}\Re(\kappa\Lambda_T^*)+3v_s^2|\kappa|^2-v_d v_u\Re(\lambda\kappa^*)+\frac{1}{2}(v_d^2+v_u^2)|\lambda|^2,\nonumber\\
	&+\sqrt{2}v_s\Re(T_\kappa),\nonumber\\
	m_{\phi_d \phi_T}&=-\frac{1}{2}(g_1^2+g_2^2)v_d v_u+\frac{1}{2}v_u v_{\bar{T}}\Re(\lambda\Lambda_T^*)-v_u
	v_s\Re(\lambda\chi_d^*)+4v_d v_T|\chi_d|^2+\sqrt{2}v_d\Re(T_{\chi_d}),\nonumber\\
\end{align}
\begin{align}
      m_{\phi_u \phi_T}&=\frac{1}{2}(g_1^2+g_2^2)v_d v_T-v_u v_s\Re(\Lambda_T\chi_t^*)-v_d
      v_s\Re(\lambda\chi_d^*)+\frac{1}{2}v_d v_{\bar{T}}\Re(\lambda\Lambda_T^*),\nonumber\\
      m_{\phi_s \phi_T}&=v_s v_T|\lambda|^2-\frac{1}{\sqrt{2}}v_{\bar{T}}\Re(T_\Lambda)-
      v_s v_{\bar{T}}\Re(\kappa\Lambda_T^*)-v_d v_u \Re(\lambda\chi_d^*)-\frac{1}{2}v_u^2\Re(\chi_t\Lambda_T^*),\nonumber\\
      m_{\phi_T,\phi_T}&=m_T^2-\frac{1}{4}(g_1^2+g_2^2)(2v_{\bar{T}}^2-6v_T^2-v_u^2+v_d^2)+2v_d^2|\chi_d|^2
      +\frac{1}{2}(v_s^2+v_{\bar{T}}^2)|\Lambda_T|^2,\nonumber\\
      m_{\phi_d,\phi_{\bar{T}}}&=\frac{1}{2}(g_1^2+g_2^2)v_d v_{\bar{T}}+\frac{1}{2}v_T v_u \Re(\lambda\Lambda_T^*)
      -v_d v_s \Re(\chi_d\Lambda_T^*)-v_s v_u\Re(\chi_t\lambda^*),\nonumber\\
      m_{\phi_u,\phi_{\bar{T}}}&=-\frac{1}{2}(g_1^2+g_2^2)v_{\bar{T}}v_u+\sqrt{2}v_u\Re(T_{\chi_t})
      +\frac{1}{2}v_d v_T\Re(\lambda\Lambda_T^*)-v_d v_s \Re(\lambda\chi_t^*)+4v_{\bar{T}}v_u|\chi_t|^2,\nonumber\\
      m_{\phi_s,\phi_{\bar{T}}}&=v_s v_{\bar{T}}|\Lambda_T|^2-\frac{1}{\sqrt{2}}v_T\Re(T_\Lambda)
      -v_s v_T\Re(\kappa\Lambda_T^*)-\frac{1}{2}v_d^2\Re(\Lambda_T \chi_d^*)-v_d v_u\Re(\lambda\chi_t^*),\nonumber\\
      m_{\phi_T,\phi_{\bar{T}}}&=-(g_1^2+g_2^2)v_T v_{\bar{T}}-\frac{1}{\sqrt{2}}v_s \Re(T_\Lambda)
      -\frac{1}{2}v_s^2 \Re(\kappa\Lambda_T^*)+\frac{1}{2}v_d v_u \Re(\lambda\Lambda_T^*)+v_T v_{\bar{T}}|\Lambda_T|^2,\nonumber\\
      m_{\phi_{\bar{T}},\phi_{\bar{T}}}&=m_{\bar{T}}^2+\frac{1}{4}(g_1^2+g_2^2)(v_d^2-v_u^2-2v_T^2+6v_{\bar{T}})+
      \frac{1}{2}(v_s^2+v_T^2)|\Lambda_T|^2+2v_u^2|\chi_t|^2.
    \end{align}

\section{Constants $C_{abc}^{L,R}$ and factors appeared in our calculation. \label{constants}}
Defining $C_{abc}^{L,R}$ in neutron, heavy quarks and mercury EDM calculation, we can find :
\begin{eqnarray}
	&&C_{\bar{\tilde g}\tilde q_i q_j}^L=-\sqrt2 g_3 Z^{\tilde q}_{i,j} e^{i\theta_3},\;\;
	C_{\bar{\tilde g}\tilde q_i q_j}^R=\sqrt2 g_3 Z^{\tilde q}_{i,j+3} e^{-i\theta_3},\nonumber\\
	&&C_{\bar{q}_j\tilde q_i \tilde g}^L=\sqrt2 g_3 Z^{\tilde q *}_{i,j+3} e^{i\theta_3},\;\;
	C_{\bar{q}_j\tilde q_i \tilde g}^R=-\sqrt2 g_3 Z^{\tilde q *}_{i,j} e^{-i\theta_3},\nonumber\\
	&&C_{\bar{\chi}_k^0 \tilde u_i u_j}^L=-\frac{1}{6}\Big(\sqrt2 (g_1 Z^{N*}_{k,1}+3 g_2 Z^{N*}_{k,2})Z^{\tilde u}_{i,j}+6Y_{u,j}Z^{N*}_{k,4}Z^{\tilde u}_{i,j+3}\Big),\nonumber\\
	&&C_{\bar{\chi}_k^0 \tilde u_i u_j}^R=-\frac{1}{6}\Big(\sqrt2(3g_2Z^{N}_{k,2} +g_1 Z^{N}_{k,1})Z^{\tilde u*}_{i,j}+6Y_{u,j}^*Z^{N}_{k,4}Z^{\tilde u*}_{i,j+3}\Big),\nonumber\\
	&&C_{\bar{\chi}_k^0 \tilde d_i d_j}^L=-\frac{1}{6}\Big(\sqrt2 (g_1 Z^{N*}_{k,1}-3 g_2 Z^{N*}_{k,2})Z^{\tilde d}_{i,j}+6Y_{d,j}Z^{N*}_{k,3}Z^{\tilde d}_{i,j+3}\Big),\nonumber\\
	&&C_{\bar{\chi}_k^0 \tilde d_i d_j}^R=-\frac{1}{6}\Big(\sqrt2(-3g_2Z^{N}_{k,2} +g_1 Z^{N}_{k,1})Z^{\tilde d*}_{i,j}+6Y_{d,j}^*Z^{N}_{k,3}Z^{\tilde d*}_{i,j+3}\Big),\nonumber\\
	&&C_{\bar d_jd \tilde u_i {\chi}_k^-}^L=U_{k,2}^*\sum_{a=1}^3Y_{d,a}Z^{CKM*}_{a,j} Z^{\tilde u*}_{i,a},\nonumber\\
	&&C_{\bar d_jd \tilde u_i {\chi}_k^-}^R=\sum_{a=1}^3 \Big(-g_2 V_{k,1}Z^{CKM}_{a,j} Z^{\tilde u*}_{i,a}+V_{k,2} Y_{u,a}^*Z^{CKM}_{a,j} Z^{\tilde u*}_{i,a+3}\Big),\nonumber\\
    &&C_{\bar{\chi}_k^- \tilde d_i u_j}^L=\sum_{a=1}^3 \Big(-g_2 U_{k,1}^*Z^{CKM*}_{j,a} Z^{\tilde d}_{i,a}+U_{k,2}^* Y_{d,a}Z^{CKM*}_{j,a} Z^{\tilde d}_{i,a+3}\Big),\nonumber\\
    &&C_{\bar{\chi}_k^- \tilde d_i u_j}^R=V_{k,2}\sum_{a=1}^3Y_{u,a}^*Z^{CKM}_{j,a} Z^{\tilde d}_{i,a},\nonumber\\
    &&C_{\bar{\chi}_k^- \tilde d_i u_j}^L=\sum_{a=1}^3 \Big(-g_2 U_{k,1}^*Z^{CKM*}_{j,a} Z^{\tilde d}_{i,a}+U_{k,2}^* Y_{d,a}Z^{CKM*}_{j,a} Z^{\tilde d}_{i,a+3}\Big),\nonumber\\
    &&C_{\bar{\chi}_k^- \tilde d_i u_j}^R=V_{k,2}\sum_{a=1}^3Y_{u,a}Z^{CKM}_{j,a} Z^{\tilde d}_{i,a},\nonumber\\
    &&C_{\bar u_j \tilde d_i {\chi}_k^+}^L=V_{k,2}^*\sum_{a=1}^3Y_{u,a}Z^{CKM}_{j,a} Z^{\tilde d}_{i,a},\nonumber\\
    &&C_{\bar u_j \tilde d_i {\chi}_k^+}^R=\sum_{a=1}^3 \Big(-g_2 U_{k,1}Z^{CKM}_{j,a} Z^{\tilde d}_{i,a}+U_{k,2} Y_{d,a}Z^{CKM}_{j,a} Z^{\tilde d}_{i,a+3}\Big),\nonumber\\
    &&C_{\bar u_i h_k u_i}=-\frac{1}{\sqrt2}Y_{u_i}Z^h_{k,2},\nonumber\\
    &&C_{\bar d_i h_k d_i}=-\frac{1}{\sqrt2}Y_{d_i}Z^h_{k,1}\nonumber\\.
	\label{15}
\end{eqnarray}
Defining $C_{abc}^{L,R}$ and ${x_i} = \frac{{m_i^2}}{{m_W^2}}$ in electron EDM calculation, we can find :
 \begin{eqnarray}
    &&C_{\bar e_i \tilde v_k \tilde {\chi}_j^-}^L=U_{j,2}^*\sum_{b=1}^3 Z_{kb}^{V,*}\sum_{a=1}^3 U_{R,ia}^{e,*} Y_{e,ab}\ ,\nonumber\\
    &&C_{\bar e_i \tilde v_k \tilde {\chi}_j^-}^R=-g_2\sum_{a=1}^3 Z_{ka}^{V,*} U_{L,ia}^e V_{j1}\ ,\nonumber\\
    &&C_{\tilde {\chi}_j^+ \tilde v_k^* e_j}^L=-g_2 V_{i1}^* \sum_{a=1}^3 U_{L,ja}^{e,*} Z_{ka}^V\ ,\nonumber\\
    &&C_{\tilde {\chi}_j^+ \tilde v_k^* e_j}^R=\sum_{b=1}^3\sum_{a=1}^3 Y_{e,ab}^* U_{R,ja}^{e} Z_{kb}^{V} U_{i2}\ ,\nonumber\\
    &&C_{\tilde {\chi}_i^0 \tilde e_k^* e_j}^L=\frac{1}{2}(-2N_{i3}^*\sum_{b=1}^3U_{L,jb}^{e,*}\sum_{a=1}^3Y_{e,ab} Z_{k3+a}^{E}+\sqrt{2}g_1N_{i1}^*\sum_{a=1}^3U_{L,ja}^{e,*}Z_{ka}^{E}\nonumber\\
    &&\qquad\qquad\;+\sqrt{2}g_2N_{i2}^*\sum_{a=1}^3U_{L,ja}^{e,*}Z_{ka}^{E})\ ,\nonumber\\
    &&C_{\tilde {\chi}_i^0 \tilde e_k^* e_j}^R=-\sqrt{2}g_1\sum_{a=1}^3Z_{k3+a}^{E}U_{R,ja}^{e}N_{i1}-\sum_{b=1}^3\sum_{a=1}^3Y_{e,ab}^*U_{R,ja}^{e}Z_{kb}^{E}N_{i3}\ ,\nonumber\\
    &&C_{\bar e_i \tilde e_k \tilde {\chi}_j^0}^L=-2N_{j3}^*\sum_{b=1}^3Z_{kb}^{E,*}\sum_{a=1}^3U_{R,ia}^{e,*}Y_{e,ab}-\sqrt{2}g_1N_{j1}^*\sum_{a=1}^3Z_{k3+a}^{E,*}U_{R,ia}^{e,*}\ ,\nonumber\\
    &&C_{\bar e_i \tilde e_k \tilde {\chi}_j^0}^R=\frac{1}{2}(-2\sum_{b=1}^3\sum_{a=1}^3Y_{e,ab}^{e,*}Z_{k3+a}^{E,*}U_{L,ib}^eN_{j3}+\sqrt{2}\sum_{a=1}^3Z_{ka}^{E,*}U_{L,ia}^e(g_1N_{j1}+g_2N_j2))\nonumber\\
\end{eqnarray}
 The functions $I_{1,2,3}(x)$ in \ref{sec3.1} and $I_{1,2,3}(x_1,x_2)$ in \ref{secC} can be written as
 \begin{eqnarray}
    &&I_1(x)=\frac{1}{2(x-1)^2}(1+x+\frac{2x}{x-1}\ln x),\nonumber\\
    &&I_2(x)=\frac{1}{6(x-1)^2}(10x-26-\frac{2x-18}{x-1}\ln x),\nonumber\\
    &&I_3(x)=\frac{1}{2(x-1)^2}(3-x+\frac{2}{x-1}\ln x),\nonumber\\
	&&{I_1}(\textit{x}_1 , x_2 ) = \frac{1}{{16{\pi ^2}}}\Big[ \frac{{1 + \ln {x_2}}}{{({x_2} - {x_1})}} + \frac{{{x_1}\ln {x_1}}-{{x_2}\ln {x_2}}}{{{{({x_2} - {x_1})}^2}}} \Big]\:,\nonumber\\
\end{eqnarray}
\begin{eqnarray}
	&&{I_2}(\textit{x}_1 , x_2 ) = \frac{1}{{16{\pi ^2}}}\Big[ - \frac{{1 + \ln {x_1}}}{{({x_2} - {x_1})}} - \frac{{{x_1}\ln {x_1}}-{{x_2}\ln {x_2}}}{{{{({x_2} - {x_1})}^2}}} \Big]\:,\nonumber\\
	&&{I_3}(\textit{x}_1 , x_2 ) = \frac{1}{{32{\pi ^2}}}\Big[  \frac{{3 + 2\ln {x_2}}}{{({x_2} - {x_1})}} - \frac{{2{x_2} + 4{x_2}\ln {x_2}}}{{{{({x_2} - {x_1})}^2}}} -\frac{{2x_1^2\ln {x_1}}}{{{{({x_2} - {x_1})}^3}}} \nonumber\\
	&&\qquad\qquad\quad\; + \: \frac{{2x_2^2\ln {x_2}}}{{{{({x_2} - {x_1})}^3}}}\Big]\:, \nonumber\\
	&&{I_4}(\textit{x}_1 , x_2 ) = \frac{1}{{96{\pi ^2}}} \Big[ \frac{{11 + 6\ln {x_2}}}{{({x_2} - {x_1})}}- \frac{{15{x_2} + 18{x_2}\ln {x_2}}}{{{{({x_2} - {x_1})}^2}}} + \frac{{6x_2^2 + 18x_2^2\ln {x_2}}}{{{{({x_2} - {x_1})}^3}}}  \nonumber\\
	&&\qquad\qquad\quad\; + \: \frac{{6x_1^3\ln {x_1}}-{6x_2^3\ln {x_2}}}{{{{({x_2} - {x_1})}^4}}}  \Big]\:.\nonumber\\
\end{eqnarray}
 \section{Effective Lagrangian in our calculation. \label{Lagrangian}}
 The $\bar q \tilde q_i g$  interaction is given by~\cite{Cao:2014kya}
\begin{equation}
	{\cal{L} }_{eff}= - \sqrt{2} g_s \sum_{a} T^a \left\{ \bar{\tilde{u}}_{L} V_{u}^{LL} u_{L} - \bar{\tilde{u}}_{R} V_{u}^{RR} u_{R} + \bar{\tilde{d}}_{L} V_{d}^{LL} d_{L} - \bar{\tilde{d}}_{R} V_{d}^{RR} d_{R} \right\}  + H.c.
\end{equation} 	
The $\bar q \tilde q_i \chi_j^+$  interaction is given by~\cite{Ibrahim:2004gb}
\begin{eqnarray}
{\cal L}_{eff}= g\bar t ((R_{bij} + \Delta R_{bij}) P_R +
(L_{bij}+\Delta L_{bij})P_L) \tilde \chi_j^+\tilde b_i\nonumber\\
+g\bar b ( (R_{tij}+ \Delta R_{tij})  P_R + (L_{tij} + \Delta L_{tij})P_L) \tilde \chi_j^c \tilde t_i +H.c.
\end{eqnarray}
The $\bar q \tilde q_i \chi_j^0$  interaction is given by~\cite{ Ibrahim:2004gb}
\begin{eqnarray}
{\cal {L}}_{eff}= g\bar b [(K_{bij} + \Delta K_{bij})   P_R +
(M_{bij} + \Delta M_{bij}) P_L]\chi_j^0 \tilde b_i\nonumber\\
+g \bar t [  (K_{tij}+\Delta  K_{tij})P_R +
(M_{tij} + \Delta M_{tij}) P_L]\chi_j^0 \tilde t_i + H.c.
\end{eqnarray}

\end{document}